\documentclass[ijoo,nonblindrev]{informs-ijoo}
\RequirePackage[OT1]{fontenc}
\RequirePackage{
amsmath,
times,
graphicx,
amsfonts,
amscd,
amssymb,
algorithm,
algorithmic,
bbm,url
}
\OneAndAHalfSpacedXI

\usepackage[sort&compress,comma,authoryear]{natbib}
 \bibpunct[, ]{(}{)}{,}{a}{}{,}%
 \def\bibfont{\small}%
 \def\bibsep{\smallskipamount}%
 \def\bibhang{24pt}%
\TheoremsNumberedThrough     
\ECRepeatTheorems

\EquationsNumberedThrough    

\MANUSCRIPTNO{} 

\newcommand{\paranth}[1]{\left(#1\right)}

\newcommand{\curly}[1]{\left\{#1\right\}}

\numberwithin{equation}{section}
\theoremstyle{plain}
\newtheorem{thm}{Theorem}[section]

\newtheorem{lem}{Lemma}

\newcommand{\E}{\mathcal{E}}

\newcommand{\hunter}[1]{\textnormal{#1}}

\begin{document}


\RUNAUTHOR{Hunter, Vielma, and Zaman}

\RUNTITLE{Picking Winners in Daily Fantasy Sports Using Integer Programming}

\TITLE{Picking Winners in Daily Fantasy Sports Using Integer Programming}

\ARTICLEAUTHORS{%
\AUTHOR{David Scott Hunter}
\AFF{Operations Research Center, Massachusetts
Institute of Technology, Cambridge, MA 02139, \EMAIL{dshunter@mit.edu}} 

\AUTHOR{Juan Pablo Vielma}
\AFF{Department of Operations Research, Sloan School of Management, Massachusetts
Institute of Technology, Cambridge, MA 02139, \EMAIL{jvielma@mit.edu}}

\AUTHOR{Tauhid Zaman}
\AFF{Department of Operations Management, Sloan School of Management, Massachusetts
Institute of Technology, Cambridge, MA 02139, \EMAIL{zlisto@mit.edu}}
} 

\ABSTRACT{

 We consider the problem of selecting a portfolio of entries of fixed cardinality for contests with top-heavy payoff structures, i.e. most of the winnings go to the top-ranked entries. This framework is general and can be used to model a variety of problems, such as movie studios selecting movies to produce, venture capital firms picking start-up companies {to invest in}, or individuals selecting {lineups} for daily fantasy sports contests, which is the example we focus on here.  We model the portfolio selection task as a combinatorial optimization problem with a submodular objective function, which is {given by} the probability of at least one entry winning.  We then show that {this probability} can be approximated using only pairwise marginal probabilities of the entries winning when there is a certain structure on their joint distribution.  We consider a model where the entries are jointly Gaussian random variables and present a closed form approximation to the objective function.  {Building on this,} we then consider a scenario where the entries are given by sums of constrained resources and present {an} integer programming formulation to construct the entries.  Our formulation uses principles based on our theoretical analysis to construct entries: {we} maximize the expected score of an entry {subject to a} lower bound on its variance and {an} upper bound {on} its correlation with previously
 constructed entries.      To demonstrate the effectiveness of our integer programming approach, we apply it to
 daily fantasy sports contests that have top-heavy payoff structures.
  We find that {our approach performs} well in practice.  Using our integer programming approach, we {are} able to rank in the {top-ten} multiple times in hockey and baseball contests with thousands of {competing entries}.   Our approach  can easily be extended to other problems with constrained resources and a top-heavy payoff structure.
}

 \HISTORY{}

\maketitle

\section{Introduction}\label{sec:intro}

Imagine one must select a portfolio of entries for a contest where all or nearly all of the winnings
go to the top performing entry.  Because of the top-heavy payoff structure, one natural strategy is to select the entries in such a way that maximizes the probability that at least one of them has exceptional performance and wins the contest. We refer to this as the \emph{picking winners} problem. In \cite{Hunter} the picking winners problem {is} introduced, and {this study illustrates how} this framework can be used to model the challenge faced by a venture capital fund. In particular, a venture capital fund will invest in a portfolio of start up companies and due to the top-heavy return structure, if one of the companies becomes a huge success, the fund will earn a large return. The picking winners framework can be also be used to model a wide variety of other problems {that} have top-heavy payoff structure. For instance, pharmaceutical companies want to select a set of drugs {to} invest research resources {in} such that the probability at least one of the drugs is a major success is maximized. For another example, consider a movie studio that has to select a set of movies to produce{,} and the movie studio achieves success if one of the movies is a blockbuster hit.    This framework can also model daily fantasy sports contests where people  enter lineups of players in a sport (such as hockey or baseball) into daily contests where the top scoring entry wins a huge amount of money. All of these examples require one to select a set of entries (drugs, start-up companies, movies, lineups of players) to maximize the probability that at least one of them has exceptional performance and wins.

Many applications of the picking winners framework require one to select entries from a fixed set, such as {start-up} companies for venture capital funds.
However, there are also scenarios where the set of entries is not provided in advance.  Instead, the entries must be constructed from a set of constrained resources.  One example of this is daily fantasy sports contests, where one must select players for a lineup, but the players must satisfy budget and position constraints  imposed by the company running the contests. For instance, in hockey contests, a lineup can only have one goalie.  In these types of situations one must develop a method to construct the entries in a manner \text{that} respects the constraints but still maximizes the probability of winning.

To select the entries one requires some predictive model of how well the entries will perform.  However, this information alone will not be sufficient to select a good set of entries.  The goal is to maximize the probability that at least one entry wins.  One may select entries that have a high probability of individually winning.  However, if these entries have a high positive correlation, then the probability that at least one of them wins may not be as large as for a set of  entries with slightly lower winning probabilities{,} but lower correlation so they cover more possible outcomes.  Therefore, choosing a good set of entries would require information on their joint distribution in addition to their marginal probabilities of winning.  Intuitively, a good set of entries will each have a high probability of winning, but also not be too positively correlated in order to diversify the set of entries.  What one needs to do is understand the {trade-off} between the winning probability of the entries and their correlations in order to select a good set {of} entries.

In \cite{Hunter} it was shown that the general picking winners problem is NP-hard. Nevertheless, it has a non-negative, non-decreasing, and submodular objective function{,} which implies that a greedy solution has strong performance guarantees. Naturally then, these results motivate using a greedy approach to solve the picking winners problem. However, \cite{Hunter} focuses on the scenario where at each iteration of the greedy approach one simply chooses the entry (for instance, a start-up company) from a particular ground set {that} maximizes the marginal gain in the objective function. In this study, we will focus on the situation where the entries must be constructed from constrained resources and cannot simply be chosen from a ground set. Our problem is closely related to the problem of maximizing a stochastic monotone submodular function with respect to a matroid constraint \citep{asadpour2015maximizing}. However in our case the objective function is the probability of at least one entry winning, which many times lacks a closed form expression. Therefore, we cannot even evaluate the objective function for a set of entries, which will prevent us from directly applying the greedy approach.

To overcome these challenges, we develop a greedy {approach} for constructing entries from constrained resources for top-heavy contests{, where at each step in the greedy approach we solve an integer programming problem}.  Our approach is fairly general and can be applied to a variety of problems.
 To demonstrate the effectiveness of our approach, we use it to construct entries for daily fantasy sports competitions.  We were able to perform exceptionally well in these contests across multiple sports.  To provide more context for our results, we next present a brief overview of the structure of daily fantasy sports contests.

\subsection{Daily Fantasy Sports Contests}\label{sec:dfs}

Daily fantasy sports have recently become a huge industry.  For instance, DraftKings, the leading company in this area, reported \$30 million in revenues and \$304 million in entry fees in 2014 \citep{ref:dfs_size}.   While there is an ongoing legal debate as to whether or not daily fantasy sports contests are gambling \hunter{\citep{ref:gambling,eastondaily,getty2018luck}}, there is a strong element of skill required to continuously win contests as evidenced by the fact that the {top-one} percent of players pay 40 percent of the entry fees and gain 91 percent of the profits \citep{ref:winners}.    This space is competitive, with many users employing advanced analytics to pick winning lineups.  Several sites such as Rotogrinders have been created {that} solely focus on daily fantasy sports analytics \citep{ref:rotogrinders}.  Other sites such as Daily Fantasy Nerd even offer to provide optimized contest lineups for a fee \citep{ref:dfn}.

There are a variety of contests in daily fantasy sports with different payoff structures.  Some contests distribute all the entry fees equally among all lineups that score above the median score of the contest.  These are known as double-up or 50/50 contests.  They are relatively easy to win, but do not pay out much.  Other contests give a disproportionately high fraction of the entry fees to the top scoring lineups.  We refer to these as \emph{top-heavy} contests.    Winning a top-heavy contest is much more difficult than winning  a double-up  or  a 50/50 contest, but the payoff is also much greater.  Typical entry fees range from \$3 to \$30, while the potential winnings range from thousands to a million dollars \citep{ref:dk}.  These contests occur nearly {every day}, providing multiple opportunities to test and evaluate different strategies.  For these reasons, daily fantasy sports provide a great opportunity for us to test our approach to
picking winners.

The entry to a daily fantasy sports contest is called a lineup{,} and {it} consists of a set of players in the given sport.   An added complication is the fact
that there are constraints on the {players} used to build a lineup.  A daily fantasy sports company sets a price (in virtual dollars) for each player in a sport. These prices vary {day-to-day} and are based upon past and predicted performance \citep{ref:dkprices}.  Lineups must then be constructed within a  fixed budget.  For instance, in DraftKings hockey contests, there is a \$50,000 budget for each lineup.  There are further constraints on the maximum number of each type of player allowed and the number of different teams represented in a lineup.  Finding an optimal lineup is challenging because the performance of each player is unknown and there are hundreds of players to choose from. However, predictions for player points can be found on a host of sites such as Rotogrinders \citep{ref:rotogrinders} or Daily Fantasy Nerd \citep{ref:dfn} {that} are fairly accurate, and many fantasy sports players have a good sense about how players will perform.  This suggests that the main challenge is not in predicting player performance, but in using predictions and other information to construct  the lineups.

{There have been many previous studies on how mathematical optimization and analytical methods can be used to assist sporting organizations. For instance,  \cite{sharp2011integer} use integer programming to construct Cricket teams, \cite{pantuso2017football} propose an optimization model to maximize the expected value of a European football club,   \cite{ozlu2016optimization} employ integer programming to solve the problem of assigning and scheduling minor league scouts for Major League Baseball teams, and \cite{sokol2003robust}  generate optimal baseball batting orders that are robust under uncertainty in skill measurement. Additionally, there have been many studies that have applied analytical methods to in consumer sporting competitions. For instance, multiple studies  have considered the problem of predicting the maximum number of winners in the National Collegiate Athletic Association (NCAA) March Madness college basketball tournament \citep{kvam2006logistic,brown2010improved,guestrin2005near,kaplan2001march}. There has also been work that considers ranking NCAA Football teams \citep{kolbush2017logistic}. }

{In addition to these studies, there has been previous research on the topic of season-long fantasy sports. For instance, in \cite{ref:footballSun} the authors present a mixed integer program that optimizes draft selection and weekly lineup selection for fantasy football.  \cite{fry2007player} present a dynamic programming approach for fantasy football draft selection. \cite{bergman2017surviving}  use optimization tools for National Football League survivor fantasy pools. All of these studies illustrate that optimization can be a useful tool in fantasy sports. However, all of these studies focus on season-long fantasy contests, which do not necessarily coincide with the problem of picking winners in daily fantasy contests that we study in this work.  }

{In a recent study  a stochastic integer programming approach is proposed to construct lineups for daily fantasy sports contests that maximizes the expected payout \citep{newell2017optimizing}. In this study, the author's assume that each athlete's fantasy points and independent. Using their approach they are able to maximize the probability of achieving a payout, but they do not find success in real DraftKings contests. We take a different approach to this problem, and with our approach we are able to win several DraftKings contests. Achieving this required  real entries in the contests that generated winnings of up to thousands of dollars per entry (See Figure~\ref{fig:dkrun}).      To avoid any related ethical or conflict of interest issues,  we donated all   winnings generated during this study (around \$15,000) to the Greater Boston Food Bank, which is a non-profit hunger-relief organization in eastern Massachusetts.}

\subsection{Our Contributions}
Integer programming is a natural tool to use to construct entries because of the different constraints.
The challenge comes in understanding how to formulate the integer program to maximize the probability of winning.
In this work we present a {sequential integer programming approach} that attempts to do this.
We begin by finding
a tractable approximation for the objective function (the probability of winning){,} and show that the error
of this approximation is small when the underlying distribution of the entries satisfies certain natural properties.  From this approximation we are able to
extract a set of heuristic principles for constructing entries.  First, the entries' scores should have a large expected value
and variance.  This increases the marginal probability of an entry winning.  Second, the entries should also have low correlation with
each other to make sure they cover a large number of possible outcomes.  These principles form the foundation of our
integer programming {approach} that sequentially constructs entries with maximal mean subject to constraints on their
variance and correlation with previously constructed lineups.

\begin{figure}
	\centering
		\includegraphics[scale=0.47]{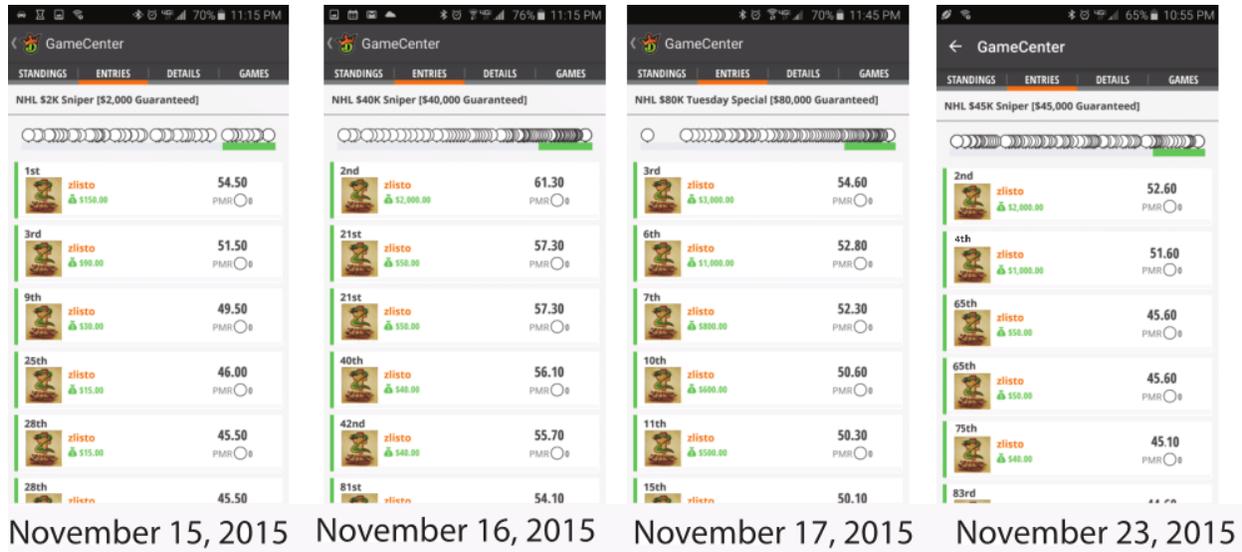}
		\caption{Screenshots of the performance of our top integer programming lineups in real DraftKings hockey contests with 200 linueps entered. }
	\label{fig:dkrun}
\end{figure}

We {apply} our approach
to top-heavy hockey and baseball contests in DraftKings.  The contests we {focus} on had thousands of entrants and top {payouts} of several thousand dollars.   The entries {that we construct} with our integer programming {approach are} able to rank in the top ten in these contests multiple times, as shown in Figure \ref{fig:dkrun}.  This is a non-trivial achievement given the large size of the contests.  {  One may be concerned that our integer programming approach is a betting tool that constitutes cheating.  However, daily fantasy sports have been shown to be games of skill \citep{getty2018luck}, so the use of such tools is considered fair game and are routinely offered as a for-profit service \citep{ref:dfn,ref:rotogrinders}.  Our hope is that by making our results and code here public, we can help to elevate the level of the players that compete by introducing them to advanced operations research and management science techniques, which they could then apply elsewhere. Nonetheless, as mentioned before, all winnings were donated to charity to avoid any conflict of interest issues.     }

{Previous studies \citep{sokol2004intuitive,kvam2004teaching} have found that sports can be used as a valuable tool to teach students analytical tools.}
Our success in using integer programming in daily fantasy sports has also exposed a whole new
audience to analytics and optimization.  We have presented our results at the Sloan
Sports Analytics Conference in 2016 and 2017 \citep{ref:ssac_hockey,ref:ssac_baseball}, giving a non-technical audience a tutorial
on integer programming via daily fantasy sports.  We have also presented our work
in multiple analytics classes for undergraduate, {graduate} and MBA {students} at MIT {and Harvard}.
{Our} optimization code {is} available on GitHub \citep{ref:github_hockey,ref:github_baseball}. Daily fantasy sports have proved itself to be an excellent application to get new people
interested in studying analytics and optimization. {In particular, graduate students at MIT routinely present extensions of this work to other sports as part of their class projects.}

{We strongly believe that a crucial aspect for the past and future success of this study as an educational tool to disseminate the effectiveness of operations research and analytics tools is its relative simplicity. Indeed, the models and tools are intuitive and interpretable for anyone with a basic background in optimization. Nonetheless, these tools provided a significant edge in an environment where  highly technical machine learning tools are widely used by both experts and users that barely understand their principles. For this reason we have purposely avoided adding complexity for complexity's sake, particularly when it does not provide a practical  benefit.}

The remainder of the paper is structured as follows.
In Section \ref{sec:problem} we reconsider the picking winners problem{,} and
develop {an algorithm to sequentially construct each entry using an integer program}.
We present an empirical analysis of hockey data
and develop predictive models of player performance in Section \ref{sec:data} for use in our integer programming
formulation.
In Section \ref{sec:algorithm} we specialize our integer programming formulation
for daily fantasy sports contests, with a focus on hockey.  We evaluate the performance of our approach
in actual fantasy hockey contests in Section \ref{sec:performance}.  We conclude in Section \ref{sec:conclusion}.  All proofs, unless otherwise stated, are included in the appendix.

 \section{Problem Formulation}\label{sec:problem}
Using similar notation to \cite{Hunter}, we consider a set of events $\mathcal E=\curly{E_i}_{i=1}^m$ defined on a probability space $(\Omega,\mathcal F,\mathbf P)$, with sample space $\Omega$, $\sigma$-algebra $\mathcal F$, and probability measure $\mathbf P$. Using this notation, each event $E_i$ corresponds to an entry $i$ in the contest winning. The goal then is to select a subset $\mathcal S \subseteq [m] = \left\{1, \ldots, m \right\}$ of size $k$ such that we maximize the probability that at least one item in $\mathcal {S}$ wins, which is given by $\mathbf{P} \left( \bigcup_{i \in \mathcal S} E_i \right)$. For notational convenience, we denote $U\left(\mathcal S\right) = \mathbf{P} \left( \bigcup_{i \in \mathcal S} E_i \right)$. Formally then, the picking winners problem is given by the following optimization problem
 \begin{align}\label{eq:opt_prob_max}
    \max_{\mathcal S\subseteq \mathcal [m], |\mathcal S|=k}~U(\mathcal S) .
 \end{align}
Before continuing with our analysis, we first begin by reviewing some of the key results from \cite{Hunter} concerning the picking winners problem. For convenience, we reproduce the proofs of these results in Section \ref{sec:proofs}. {For notation, we will let $\mathcal S_g$ be the greedy maximizer of $U(\mathcal S)$ and let $\mathcal S^*$ be the maximizer of $U(\mathcal S)$.}

{
In general, solving the picking winners optimization problem can be difficult, as shown by the following result.
\begin{thm}[\cite{Hunter}]\label{thm:nphard}
Maximizing $U(\mathcal S)$ is NP-hard even for discrete uniform distributions.
\end{thm}
Despite the computational complexity in maximizing $U(\mathcal S)$, the objective function
possesses some nice properties {that} we can use to obtain good solutions efficiently, which
are given by the following  result.
\begin{lem}[\cite{Hunter}]\label{lem:submodular}
The function $U(\mathcal S)$ is non-negative, non-decreasing and submodular.
\end{lem}
From Theorem \ref{thm:nphard}, we know that the picking winners problem given by  \eqref{eq:opt_prob_max} can be difficult. However, the submodular property from Lemma \ref{lem:submodular} allows us to find a good solution using a greedy approach where we choose the events one at a time to give the maximal increase in the objective function.  For submodular functions we have guarantees on the sub-optimality of such greedy solutions, as the following classic result shows.}

\begin{thm}[\cite{nemhauser1978analysis}] \label{thm:submodular}
{Let $U(\mathcal S)$ be a  non-decreasing submodular function and let $\mathcal S_g$ be its greedy maximizer.  Then}
\begin{align}
       \frac{U(\mathcal S_g)}{U(\mathcal S^*)}&\geq 1-e^{-1}.
\end{align}
\end{thm}
{
Theorem \ref{thm:submodular} provides a strong bound for the sub-optimality of a greedy solution to the picking winners problem. In practice, a greedy solution can often be found efficiently, which is important as time can be a limiting factor in many applications. For these reasons, we will continue by considering a greedy approach to the picking winners problem. }

Ideally, a greedy solution to the picking winners problem would also be an optimal solution. In the case of all the events $E_i$ being independent, from \cite{Hunter} we know that a greedy solution is in fact optimal:
\begin{thm}[\cite{Hunter}]\label{thm:greedy_independent}
For a set of independent events $\mathcal E = \left\{E_1, \ldots, E_m \right\}$, let the $k$ element greedy maximizer of $U ( \mathcal S )$ be \hunter{$\mathcal S_g$}. Then \hunter{$\mathcal S_g$} is an optimal solution to the picking winners problem \hunter{\eqref{eq:opt_prob_max}}.
\end{thm}
This result also suggests that when the events have low dependence, greedy solutions should perform well. In practice, the events will often have some dependency structure, and thus we will consider this case in more detail.

 \subsection{Dependent Events}
 When the underlying events are dependent, solving the picking winners problem is especially difficult. In general, a greedy solution is not an optimal solution. Additionally, with an arbitrary dependency structure between the events it is often non-trivial to even evaluate the objective function $U(\mathcal S)$. Because of this, in this section we provide a tractable approximation to the objective function when the events are dependent and characterize the quality of the approximation as a function of properties of the underlying distribution.  We will see that when the events are not too dependent upon each other, our approximation will have low error.

We start with an approximation for the objective function $U(\mathcal S)$.
The difficulty of optimizing this objective function comes from the lack of independence in the events{,} which requires us to understand their entire joint distribution.  Many times this distribution will not be available to us, or if it is, evaluating the required probability will be difficult.  To obtain something more tractable, we define the following objective function{,} which only utilizes pairwise correlations:
\begin{align}
	U_2(\mathcal S) & = \sum_{i\in \mathcal S}\mathbf P\paranth{E_i}-\frac{1}{2}\sum_{i,j\in \mathcal S,i\neq j}\mathbf P\paranth{E_i\bigcap E_j}.
\end{align}
This objective function is obtained by expanding $U(\mathcal S)$ using the inclusion-exclusion principle and then only keeping terms {that} involve the intersection of two or fewer events.  The benefit of $U_2(\mathcal S)$ is that it only requires us to know the probability of intersections of pairs of events, which is much easier to calculate than probabilities of higher order intersections.  In order for this objective function to act as a good surrogate for $U(\mathcal S)$, we must show that the difference
$U(\mathcal S)-U_2(\mathcal S)$ is small.  If we impose certain conditions on the joint density of the events, we can bound this difference.  To do this we define a few new terms.

We assume we have chosen a subset $\mathcal S$ of the entries.  Let \hunter{$q_0(\mathcal S) = \mathbf P(\bigcap_{i\in \mathcal S} E_i^c)$} be the probability that none of the entries in $\mathcal S$ win the contest. For $1\leq l \leq k$, let $\mathcal S_l$ be the set of all $l$-element subsets of $\mathcal S$.  For instance, if $\mathcal S=\curly{a,b,c}$ then $\mathcal S_1=\curly{\curly{a},\curly{b},\curly{c}}$ and $\mathcal S_2 = \curly{\curly{a,b},\curly{a,c},\curly{b,c}}$.  For an element $T\in\mathcal  S_l$, let $p'_T = \mathbf P\paranth{\paranth{\bigcap_{i\in T} E_i} \bigcap \paranth{\bigcap_{j\in \mathcal S/T} E_j^c} }$ be the probability that all the entries in $T$ but no other entries in $\mathcal S$ win the contest.  We then have the following result regarding the difference between $U(\mathcal S)$ and $U_2(\mathcal S)$.
\begin{thm}\label{thm:U2}
\hunter{
Let $\mathcal E$ be a set of events on a probability space and assume $|\mathcal E|=m$.  Assume there are constants $0<p<1$ and $c>0$ such that for any $\mathcal S\subset [m]$ with $|\mathcal S|=k<m$, $q_0(\mathcal S) = (1-p)^k$, $kp<1/2$, and for any $1\leq l \leq k$ and $T\in S_l$, $p'_T \leq cp^l$.  Then
\begin{align*}
	kp-\frac{1}{2}(kp)^2 \leq U(\mathcal S) \leq kp
\end{align*}
and
\begin{align*}
	0\leq U(\mathcal S)-U_2(\mathcal S)& \leq 20c(kp)^3.
\end{align*}
}
\end{thm}

{This result shows that the error of $U_2(\mathcal S)$ is approximately $(U(S))^3$ multiplied by a constant.  We will generally be in a situation where the events are rare, so $U(S)$ is small.  Then the cubic term in the error is negligible compared to $U(S)$.  The error then depends mainly on the constant $c${,} which is related to how correlated the events are.  We now illustrate this with an example.}

{Consider the case where $k = 3$, and assume the marginal probability of any single event occurring is $0.01$.  If the events are  independent and identically distributed (i.i.d.), then $q_0(\mathcal S) = (1-p)^3 = (1-.01)^3$, resulting in $p=0.01$. The condition $p'_T\leq cp^l$ from the theorem is satisfied with $c=1$.   For this case, we have $U(\mathcal S) = 0.029701$ and $U_2(\mathcal S) = 0.029700$, resulting in an error of $U(\mathcal S)-U_2(\mathcal S) = 10^{-6}$.  From the theorem the error is upper bounded by $20c(kp)^3 = 5.4\times 10^{-4}$.  }

{
Now consider the case where the three events are the same.  This represents a situation of maximal correlation. We have $q_0(\mathcal S) = (1-p)^3 = (1-.01)$, resulting in $p = 0.0033$.  To satisfy the condition in the theorem, we only need to consider the set $T = \mathcal S$, since for all other subsets $T$ we have $p'_T = 0$.  When $T=\mathcal S$, the condition from the theorem gives $0.01 \leq c(0.0033)^3$.  This is satisfied with $c = 2.7\times 10^{5}$.   For this case, we have $U(\mathcal S) = 0.01$ and $U_2(\mathcal S) = 0$, resulting in an error of $U(\mathcal S)-U_2(\mathcal S) = 0.01$.  From the theorem the error is upper bounded by $20c(kp)^3 = 5.4$.  We see that when the events are highly correlated, the constant $c$ increases and the upper bound becomes loose (trivial in this case).
}

{
These examples show that the error bound from the theorem is tighter when the events are not highly correlated, and the constant $c$ is a way to capture the strength of the correlation.  In the situations we consider, the events will have a low occurrence probability and will not have very strong correlations.  Therefore, we will use $U_2(\mathcal S)$ instead of $U(\mathcal S)$ in the remainder of this work because $U_2(\mathcal S)$ is much easier to work with than $U(\mathcal S)$, and it will be a good approximation to $U(\mathcal S)$.
}

\subsection{Multivariate Gaussian Random Variables}\label{sec:gaussian}
We now consider a probability model that captures a fairly general class of events.
Let $\curly{X_i}_{i=1}^m$ be jointly Gaussian random variables with mean vector $\mu$ and covariance matrix $\Sigma$.  We define the events $E_i = \curly{X_i\geq t}$ for some constant $t$.  If we imagine each $X_i$ as the return of a company we invest in, then the event says that company $i$ has a return greater than $t$.  For fantasy sports competitions, $X_i$ can be the points of lineup $i$ and $E_i$ corresponds to the lineup exceeding $t$ points.  We further assume that $t> \max_{1\leq i \leq m}\mu_i$ because we are interested in the unlikely event that one of the $X_i$ {exceeds its} mean by a large amount.  For instance, in daily fantasy hockey contests, the maximum mean of a lineup is around 30 points, while the winning threshold $t$ is near 60 points.

We want to choose a set of entries $\mathcal S\subseteq [m]$ to maximize $U_2(\mathcal S)$.  To do this, we must first calculate the marginal and joint probabilities of the events.  Since no closed form expression exists for these probabilities for multivariate Gaussian random variables, we instead resort to approximations.  We have the following result.

\begin{thm}\label{lem:lower_bound}
Let $\curly{X_i}_{i=1}^m$ be jointly Gaussian random variables.  Let the marginal mean and variance of $X_i$ be $\mu_i$ and $\sigma^2_i$ and let the correlation coefficient of $X_i$ and $X_j$ be $\rho_{ij}$.  In addition, let $z_i=(t-\mu_i)/\sigma_i$.  Define the function $U^l_2(\mathcal S)$ as
\begin{align}
	U^l_2(\mathcal S) =& \sum_{i:E_i\in \mathcal S} \frac{1}{\sqrt{2\pi}\sigma_i(z_i+1/z_i)}\exp{\paranth{-\frac{z_i^2}{2}}}\nonumber\\
	&-\frac{1}{2}\sum_{\substack{i,j:\\E_i,E_j\in \mathcal S\\i\neq j}} \frac{1}{\sqrt{2\pi}(2t-\mu_1-\mu_2)} \exp\paranth{-\frac{(2t-\mu_i-\mu_j)^2}{2\paranth{\sigma_i^2+\sigma_j^2+2\rho_{ij}\sigma_i\sigma_j}}} \label{eq:u2l}.
\end{align}
Then  $U_2(\mathcal S)\geq U_2^l(\mathcal S)$.
\end{thm}
The function $U^l_2(\mathcal S)$ is a lower bound on $U_2(\mathcal S)$ and has a closed form expression.  We use this function
to gain insights on good entries.  From the expression for $U^l_2(\mathcal S)$ we observe some important features that characterize a subset $\mathcal S$ {that} will have a high chance of winning.  We assume that $z_i\geq 1$,{,} which  can be done by having $t$ be large enough.  In this case,
the contribution of the first summation to $U^l_2(\mathcal S)$ increases for large $\mu_i$ and large $\sigma_i$.  The contribution of the second summation increases for small $\rho_{ij}$.
There is an intuitive explanation for how $\mu_i$, $\sigma_i$, and $\rho_{ij}$ impact the objective function.  First, the means of the chosen entries should be large so that there is a greater probability that they exceed $t$.  Second, the variance of the entries should be large as well.  This is less obvious, but the intuition here is that all one cares about is having some entry's score exceed $t$.  If entries with large variances are chosen, this increases the probability of exceeding $t$.  While this also increases the probability that some of the entries will have a low value, this is not a concern as the goal is to have at least one exceed $t$.  Third, the correlation between each pair of entries should be small.  What this does is diversify the selected entries so there is a better chance that one of them exceeds $t$.  In short, one would like $\mathcal S$ to consist of entries with high means and high variance {that} have low correlation with each other.

\subsection{Picking Winners with Constrained Resources}\label{sec:resources}
We now consider a case where entries are constructed from a set of resources \hunter{$\curly{Y_j}_{j=1}^p$} {that} are jointly Gaussian random variables.  We define the mean and standard deviation of \hunter{$Y_j$} as \hunter{$\mu'_j$} and \hunter{$\sigma_j'$}, respectively.  We define the correlation coefficient of $Y_j$ and $Y_l$ as $\rho_{jl}'$.  Each entry is given by the sum of $n$ resources.  Let us define $x_{ij}$ as variables {that} are one if resource \hunter{$j$} is in entry $i$ and zero otherwise.  Then we can write \hunter{$X_i = \sum_{j=1}^pY_jx_{ij}$}.  We assume that there are certain constraints on the resources such that not every possible subset of $q$ resources is a feasible entry.  In this scenario, we want to construct \hunter{$k$} feasible entries from the given resources to maximize the probability of winning.  To do this we will develop an integer programming formulation {that} builds upon the intuitions gained in Section \ref{sec:gaussian}.
{Specifically, our general paradigm for the formulation will make use of the following three insights.}
\begin{itemize}
    \item[1.] {Each entry has a high mean in order to increase the probability it exceeds the threshold $t$.}

    \item[2.] {Each entry has a high variance in order to increase the probability it exceeds the threshold $t$.}

    \item[3.] {Each pair of entries have minimal correlation to increase the probability that at least one of the entries exceeds the threshold $t$.}
\end{itemize}
 We can write the mean of  $X_i$ as \hunter{
\begin{align}
	\mu_i & = \sum_{j=1}^p\mu_j'x_{ij}.
\end{align}}
The variance of $X_i$ is \hunter{
\begin{align}
	\sigma_i^2 & = \sum_{j=1}^{p}(\sigma_j')^2x_{ij}+\sum_{j=1}^p \sum_{l=1,l\neq j}^p\rho'_{jl}\sigma_j'\sigma_l'x_{ij}x_{il},
\end{align}}
and the covariance of $X_i$ and \hunter{$X_q$} is \hunter{
\begin{align}
	\text{Cov}(X_i,X_q) & = \sum_{j=1}^{p}(\sigma_j')^2x_{ij}x_{qj}+ \sum_{j=1}^p \sum_{l=1,l\neq j}^p\rho'_{jl}\sigma_j'\sigma_l'x_{ij}x_{ql},
\end{align}}
With these expressions we want to understand how to construct a set of entries that will have a high probability of winning.
To achieve a high mean, each entry should be chosen to maximize the sum of its resources' means.  This insight is fairly obvious and natural.  The more interesting insights come from the variances and covariances of the entries.  To obtain a more intuitive understanding of what entries have good properties with respect to their variance and covariance, {suppose} the variance of each resource {is} equal to one{,} and the correlation coefficient to be either zero, $\delta$, or $-\delta$ for some $\delta>0$.  We can then characterize the variance of an entry by the number of pairs of resources that have a positive or negative correlation{,} and the covariance of a pair of entries by the number of their shared resources and
the number of pairs of resources with positive and negative correlations.  Let us define for $X_i$ the number of positively and negatively pairs of correlated resources as $n_i^+$ and $n_i^-$.  Let us also define for  $X_i$ and $X_q$ the number of positively and negatively pairs of correlated resources as $n_{iq}^+$ and $n_{iq}^-${,} and the number of shared resources as $n_{iq}$.  We will also refer to $n_{iq}$ as the \emph{overlap} of entries $X_i$ and $X_j$ because this is the number of resources they have in common.  With this notation, the variances and covariances can be written as
\begin{align}
	\sigma_i^2 & = n+\delta\paranth{n_i^+-n_i^-}\\
	\text{Cov}(X_i,X_q)& = n_{iq}+\delta\paranth{n_{iq}^+-n_{iq}^-}
\end{align}
From these expressions we gain some useful insights concerning the variances and covariances.  A high variance can be achieved if the entry has a large $n_{i}^+$ and a small $n_{i}^-$.  That is, the entry's resources should be chosen so many of them are positively correlated{,} and few of them are negatively correlated.  By avoiding negative correlations, we are making it more likely that either the resources all have a high value or all have a low value.  The covariance of a pair of entries can be reduced in two ways.  First, reduce the overlap \hunter{$n_{iq}$}.  If the entries share fewer resources, they will clearly be less correlated.  Second, increase the number of pairs of resources between the lineups that have a negative correlation{,} and decrease the number of pairs of resources with positive correlation.  Limiting the number of positively correlated resources between lineups reduces their covariance and subsequent correlation coefficient.

\subsection{Sequential Integer Programming Approach with Constrained Resources}\label{sec:greedyIP}
Constructing a set of entries that optimizes the  $U_2^l(\mathcal S)$ is non-trivial.  The function is non-linear and non-convex.  Therefore, we will not directly maximize $U_2^l(\mathcal S)$.  Instead, we will use the three properties discussed  about $U_2^l(\mathcal S)$ to guide the development of a {sequential integer programming approach} for constructing a set of entries from constrained resources {that} have a high probability of winning.  We choose a greedy approach for three important reasons.  First,
from the submodular property of the original objective function we have a lower bound on the performance of a greedy solution.
The objective function we will use in our integer programming formulation is different from the original objective function, but shares some of its
key properties, so we assume the loss in optimality with our greedy approach will not be significant.  Second, the integer programming {approach we propose}
is linear and can be solved quickly, which will be important in applications we consider later.  Third, and most importantly, we will show in Section \ref{sec:performance}  that this approach performs well on actual data.  In particular, we will see  that with this approach we are able to consistently perform well in daily fantasy sports contests.

We consider the situation described in Section \ref{sec:resources} where we want to construct $k$ entries consisting of \hunter{$n$} resources each from a set of $p$ resources.  As previously stated, our decision variable is \hunter{$x_{ij}$}{,} which is 1 if resource \hunter{$j$} is in entry $i$.  Let \hunter{$\curly{x_{ij}^*}_{j=1}^p$} correspond to the $i$th greedy entry.  This is the solution to our integer program in the $i$th iteration.  We will construct the entries sequentially in the following manner.   We choose a set of positive constants \hunter{$\curly{\epsilon_i}_{i=1}^k$} and \hunter{$\curly{\gamma_{iq}}_{i,q=1}^k$}.  Assume we have obtained the first $i-1$ entries.   We then construct the $i$th entry to maximize its mean subject to its variance being larger than $\epsilon_i${,} and its covariance with entry $q$ being less than $\gamma_{iq}$ for \hunter{$q=1, \ldots, i-1$}.    For \hunter{$i =2, \ldots, k$}, our integer program for the $i$th entry is
\begin{equation}\label{eq:ip_general}
\begin{aligned}
& \underset{x}{\text{maximize}} & & \sum_{j=1}^{p}\mu'_jx_{ij} &\\
& \text{subject to} & &\sum_{j=1}^{p}(\sigma_j')^2x_{ij}+\sum_{j=1}^p \sum_{l=1,l\neq j}^p\rho'_{jl}\sigma_j'\sigma_l'x_{ij}x_{il}\geq \epsilon_i&\\
&  & &\sum_{j=1}^{p}(\sigma_j')^2x_{ij}x^*_{qk}+ \sum_{j=1}^p \sum_{l=1,l\neq j}^p\rho'_{jl}\sigma_j'\sigma_l'x_{ij}x^*_{ql}\leq \gamma_{iq}, &\hunter{$q=1, \ldots, i-1$}\\
&&& \sum_{j=1}^{p}x_{ij}= n \\
&&& \hunter{$x \in \mathcal{G}$}.
\end{aligned}
\end{equation}
\hunter{where $\mathcal{G}\subset \left\{0, 1 \right\}^p$ is some general feasible region. For constructing the first entry we solve the integer program \eqref{eq:ip_general}, but we omit the covariance constraint from the problem.}

The greedy approach only constrains the covariance of a lineup with the previous lineups, whereas a non-greedy approach would limit all pairwise covariances.  By only constraining the covariance with previous lineups, the number of {covariance} constraints is linear in the number of previous entries.  The variance constraints are non-linear in this formulation.  We can introduce auxiliary variables to make this formulation linear as is standard in many integer programming formulations.  However, we take a different approach when we apply this approach to fantasy sports contests in Section \ref{sec:algorithm}.  We will make certain assumptions on the underlying distribution of the resources{,}  which allow us to makes these constraints linear. We will see that this allows the integer program to be solved quickly{,} and {it} still {performs} well in real applications.

Our {approach} gives one the flexibility to choose the parameters $\epsilon_i$ and $\gamma_{iq}$.  We may want the first few entries' covariances to be lower in order to force exploration of the space of entries, while for later entries this constraint can be relaxed to avoid excluding entries with large means.  However, we will see that using a single value for $\gamma_{iq}$ and for $\epsilon_i$ works well in practice. In addition, we will also see that the choice of the value for $\gamma_{iq}$ depends upon the size of the space of resources.

\section{Hockey Data Analysis}\label{sec:data}
To demonstrate the effectiveness of our {approach} for picking winners,
we apply it to top-heavy daily fantasy sports hockey contests as described in Section \ref{sec:intro}.  The entries into these contests are known as lineups{,} and {these lineups} consist of resources {that} are players in the National Hockey League (NHL), which is the main hockey league in North America.  The daily fantasy contests we focus on are from the site DraftKings.  We begin by describing the basic elements of hockey and hockey daily fantasy sports contests.  To apply our {approach}, we need to know the mean and correlation of the resources, which are the hockey players.  In this section we present models and analysis for these inputs.  Section \ref{sec:prediction} presents predictive models for players points using publicly available data{,} and Section \ref{sec:correlations} studies the structural correlations between hockey players. While our analysis focuses on DraftKings, we note that other daily fantasy sports sites have similar contests and point scoring systems.  All data in this section was collected between October 21, 2015 and December 31, 2015.
\begin{figure}
    \centering
        \includegraphics[scale=1]{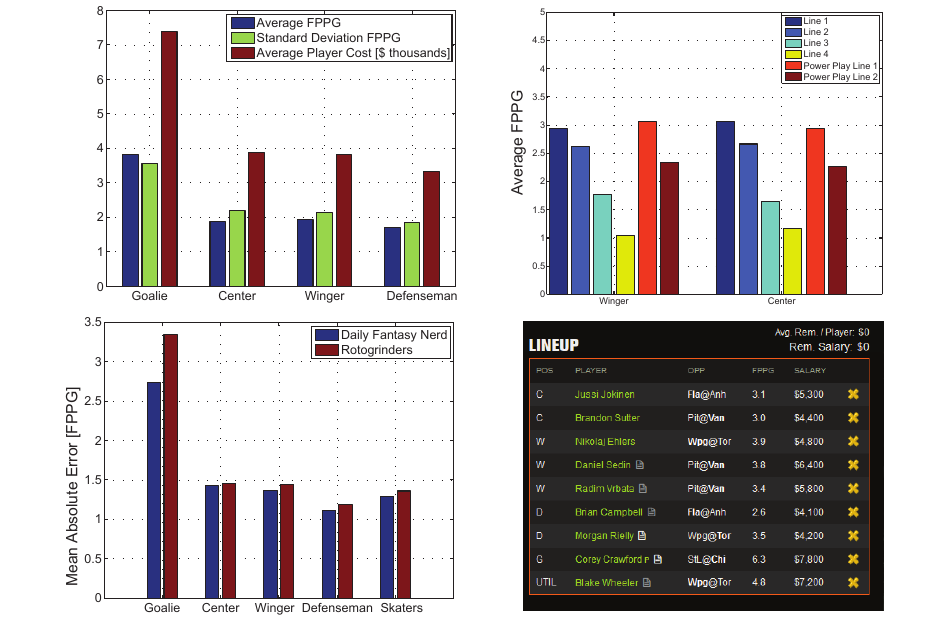}
        \caption{(top left) Mean and standard deviation of fantasy points per game (FPPG)
        and average player cost in DraftKings for each hockey position.  (bottom left) Mean absolute error of NHL player FPPG predictions of Daily Fantasy Nerd and Rotogrinders for each hockey position.
        (top right) Mean FFPG of {forwards} in different types of lines.
        (bottom right) Screenhsot of an NHL lineup from DraftKings.}
    \label{fig:figure_points_line_error}
\end{figure}

\subsection{Hockey Basics}
A hockey team consists of six players: one center, two wingers, two defensemen, and a goalie.  The centers, wingers, and defensemen are known as skaters because they are free to skate on the ice rink, whereas the goalie primarily stays in front of the goal.  In hockey the aim is for a team to shoot a puck into the goal of the opposing team and the goalie's task is to protect the goal from any opponent scores.

In daily fantasy sports hockey contests there are constraints on the types of players that must be present in a lineup.  For DraftKings, each lineup must consist of two centers, three wingers, two defensemen, one goalie, and one utility player {that} is any skater, resulting in a lineup of nine players.   We show a screenshot of one possible DraftKings lineup in Figure \ref{fig:figure_points_line_error}.  The figure also shows a column labeled FPPG, which stands for fantasy points per game.  This is the average number of fantasy points per game the player has earned throughout the current NHL season.  DraftKings provides this information to help users construct their lineups.  The total fantasy points for a lineup is the sum of each player's fantasy points.  The lineup shown has 34.4 fantasy points per game.  This score is in the typical range of NHL lineups in DraftKings{,} and {it} gives a sense of the value of different scoring actions.  In a daily fantasy sports contest, there {are many ways for players to score fantasy points}.  Because we test our approach in DraftKings, we will focus on its points scoring system.  We summarize the details of the DraftKings scoring system in Table \ref{tab:pointssystem}.  In Figure \ref{fig:figure_points_line_error} we show the basic statistics of points per game for each position averaged over all NHL players.  We also show the average DraftKings cost per player in the figure.  As can be seen, on average goalies tend to score almost twice as many points as all other players{,} and also cost twice as much.  Additionally, the standard deviation of each position's points per game is roughly the same size as the average value.

The centers and wingers are known as {forwards}.  In the NHL each hockey team plays its {forwards} in fixed sets known as \emph{lines}.  This means that if one player from the line, such as the center, is on the ice, then so are the other two players from the line.  Each team has four standard lines and two power play lines, which only play when there is a power play as a result of a penalty. We show the average points per game for players based upon their line in Figure \ref{fig:figure_points_line_error}. The rank of the line typically suggests how well it performs.  As can be seen, line one is generally the highest scoring line.

\begin{table}
    \centering
        \begin{tabular}{|l|c|}
        \hline
            Score type& Points\\ \hline
            Goal & 3\\ \hline
            Assist & 2\\ \hline
            Shot on Goal & 0.5\\ \hline
            Blocked Shot & 0.5\\ \hline
            Short Handed Point Bonus (Goal/Assist) & 1\\ \hline
            Shootout Goal & 0.2\\ \hline
            Hat Trick Bonus & 1.5\\ \hline
            Win (goalie only) & 3 \\ \hline
        Save (goalie only) & 0.2 \\ \hline
            Goal allowed (goalie only) & -1 \\ \hline
            Shutout Bonus (goalie only) & 2 \\ \hline

        \end{tabular}
    \caption{Points system for NHL contests in DraftKings.}
    \label{tab:pointssystem}
\end{table}

The points system of DraftKings is similar to most daily fantasy hockey points systems.  It gives more points for players that perform better, but it also has some important properties {that} we use to construct our lineups. The first property concerns assists, which is passing the puck to a player who either scores or who passes the puck to the player that scores.  Each assist is worth two points{,} and  DraftKings awards points for up to two assists per goal.  This is an important aspect of the scoring system because it means that for each goal scored, it is possible for a single lineup to receive seven points (three points for the goal, and four points for the two assists that led to the goal).  However, such a situation can only occur if all players involved are on a single lineup.  This suggests that to score many points, a lineup should contain all players from a single line.  By putting all players from a single line in a lineup, we are essentially increasing the variance of the lineup.  This is a popular fantasy sports tactic known as \emph{stacking}.  Many expert daily fantasy sports players use stacking to win in a variety of sports contests \hunter{\citep{ref:stacking_baseball,ref:stacking_hockey,ref:stacking_football}}. The second property of the points system concerns goalies.  If a goalie's team wins the game, an additional three points is awarded to the goalie.  This is important because as we will see, it is possible to predict who will win an NHL hockey game with relatively good accuracy, so by choosing goalies on winning teams, an extra three points can be obtained, which is a significant amount in hockey contests given that  lineups average between 30 to 40 points.

\subsection{Player Point Predictions}\label{sec:prediction}

Many fantasy sports players devote a substantial amount of effort in developing elaborate prediction models for players' performance.  Their advantage comes mainly from the
power of their predictive models. However, we do not have the domain knowledge required to improve upon the prediction models of the top fantasy players.  Therefore,
we take a different approach.  Publicly available player point predictions are provided by the websites Rotogrinders \citep{ref:rotogrinders} and Daily Fantasy Nerd \citep{ref:dfn}.  We find the predictions from these sites to be relatively accurate, but the accuracy varies by position.  We plot in Figure \ref{fig:figure_points_line_error} the mean absolute error in points of the two sites' predictions for different hockey positions using  our data from the 2015 NHL season. {We obtain each of our predictions from Rotogrinders and Daily Fantasy Nerd roughly thirty minutes before the corresponding DraftKings contest began.}  We see that the prediction errors of both sites are similar.  Therefore, it does not seem that either one has a superior prediction model.

We build a predictive model for the fantasy points for each player in the NHL using these prediction sites.
For a player $k$, let $f_{k1}$ and $f_{k2}$ be the predictions by Rotogrinders and Daily Fantasy Nerd, respectively.  For skaters, we perform a linear regression on the player's points using these predictions as features.  We denote $\mu_k$ as the predicted points of player $k$ using our model.  For skater $k$ our prediction  is
\begin{align}
\mu_k = \beta_0 + \beta_1 f_{1k} + \beta_2 f_{2k} + \hunter{$\epsilon_k$}
\end{align}
\hunter{where $\epsilon_k$ is a Gaussian error term.} For goalies, we must also factor in the outcome of the game since there is a three point bonus for victory. Daily Fantasy Nerd provides win probabilities for each goalie, and we have found these estimates to be relatively accurate. We perform a linear regression on the goalie points using as predictor variables the predictions of Rotogrinders, Daily Fantasy Nerd, and the win probability of the goalie's team. For a goalie $k$, let this win probability by $p_k$. Then the prediction for goalie $k$ is
\begin{align}
\mu_k = \beta_0 + \beta_1f_{1k} + \beta_2f_{2k} + \beta_3 p_k + \hunter{$\epsilon_k$}
\end{align}
\hunter{where $\epsilon_k$ is a Gaussian error term.} We compare different prediction models in Table \ref{table:regression}.  For skaters, both Rotogrinders and Daily Fantasy Nerd predictions are significant.   For goalies,
something different occurs. When we do not include the win probability we find that only Rotogrinders is significant.  When the win probability is included, it is the only significant feature.  {From Table \ref{table:regression} we see that the models do not have high accuracy.  However, our metric of interest is not accuracy, but performance in  real hockey contests.  We will test these different models in these contests in Section \ref{sec:performance}.}

\def\sym#1{\ifmmode^{#1}\else\(^{#1}\)\fi}
\begin{table}
\begin{tabular}{{l|}*6{c}}
\hline\hline
            &\multicolumn{1}{c}{(1)}&\multicolumn{1}{c}{(2)}&\multicolumn{1}{c}{(3)}&\multicolumn{1}{c}{(4)}&\multicolumn{1}{c}{(5)}&\multicolumn{1}{c}{(6)}\\
            &\multicolumn{1}{c}{Skater points}&\multicolumn{1}{c}{Goalie points}&\multicolumn{1}{c}{Goalie points}&\multicolumn{1}{c}{Goalie points}&\multicolumn{1}{c}{Goalie points}&\multicolumn{1}{c}{Goalie points}\\
\hline
$\beta_1$        & 0.634\sym{***}& 0.628\sym{***}&   0.203     &      & 0.203 &  \\
(Rotogrinders)   & (0.0203)      &(0.0864)       & (0.144)  &      & (0.144)  & \\
[1em]
$\beta_2$            & 0.282\sym{***}&  -0.0173 & -0.0309 &-0.0301 &  & \\
(Daily Fantasy Nerd) &    (0.0242)  &  (0.106)  & (0.113)   & (0.113)  &      &     \\
[1em]
$\beta_3$  &   &    & 4.310\sym{*}  & 5.304\sym{**} & 4.176\sym{*}  &  5.172\sym{**} \\
(Win probability) &    &   & (2.123) & (2.005) & (2.064) & (1.941) \\
[1em]
$\beta_0$&  1.334   & 1.686 \\
         & (0.0314) & (0.549) & (1.032) & (1.003) & (1.013) & (0.983)\\
\hline
$R^2$   & 0.238   & 0.0858 & 0.0179 & 0.0140  & 0.0178 & 0.0139 \\
Samples     &   10825 & 565  & 506   & 506 & 506 & 506 \\
\hline\hline
\multicolumn{7}{l}{\footnotesize Standard errors in parentheses}\\
\multicolumn{7}{l}{\footnotesize \sym{*} \(p<0.05\), \sym{**} \(p<0.01\), \sym{***} \(p<0.001\)}\\
\end{tabular}
\caption{Regression coefficients, R squared, and sample count for different prediction models of skater and goalie points.}\label{table:regression}
\end{table}


\subsection{Player Correlations}\label{sec:correlations}

\begin{figure}
    \centering
        \includegraphics[scale=1]{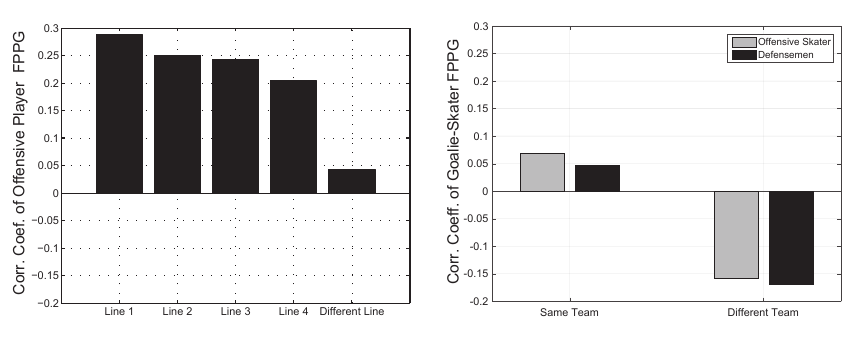}
        \caption{(left) Average correlation coefficient of FPPG for {forwards} in the same and different lines.  (right) Average correlation coefficient of FPPG for goalies and skaters in the same and opposing teams.}
    \label{fig:correlation}
\end{figure}

To create high variance lineups we must first understand the correlations between players.  There are two important correlations we look at.  The first concerns players in lines.  The second concerns {forwards} and goalies.

  We first begin by examining the correlations of FPPG between players in the same line versus those in different lines.  We  calculate the Pearson correlation coefficient between the FPPG for every pair of NHL {forwards} (centers and wingers) in the same line{,} and {on} the same team but different line.  We plot the resulting average correlation coefficients for each group in Figure \ref{fig:correlation}.  We see that the players on the same line have a much higher correlation than players on different lines.  This is not surprising given the way fantasy points are awarded.  This correlation is most likely due to the assist/goal combinations that give points to multiple players who are simultaneously on the ice.  This correlation should be included in a lineup to increase the variance.

	We next look at the correlation between goalies and skaters in the same game.  We consider the correlation of FPPG for {forwards} and defensemen with goalies  on the same and opposing team in a game.  
	We plot the resulting average Pearson correlation coefficient of the FPPG for these groups in Figure \ref{fig:correlation}.    We find that the correlation is negative for goalies and skaters on opposing teams.  For the same team the correlation is near zero.  The negative correlation is due to the fact that if a skater scores fantasy points, it is typically because a goal was scored, which gives negative fantasy points to the goalie.  On the same team the only structural condition that would correlate skaters and goalies is the win bonus and points for scoring goals.  However, this is  much weaker than the negative correlation of goalies opposing skaters.  The negative correlation is something we want to avoid within a lineup as it would reduce the overall variance.  Therefore, to increase lineup variance, we will avoid putting goalies and skaters on opposing teams in a single lineup.


\section{Fantasy Hockey {Approach and} Integer Programming Formulations}\label{sec:algorithm}
We now present our {approach} for selecting multiple lineups for hockey contests.
The integer {program} is based upon the formulation in Section \ref{sec:greedyIP}.  For the mean points of a player
we {use} our predictive regression models.  Due to difficulties in estimation, we do not explicitly use the correlations and variances of the player's points.
Instead we assume that the players all have the same variance and that the magnitude of the covariance is much smaller
than the variance.  This will lead to simple constraints for the variance and covariance.  We will construct $m$ lineups from a set of $p$ NHL players{,} and we let the variable $x_{ik}$ equal one if player $k$ is selected for  lineup $i${,} and zero otherwise.  Each player $k$ is predicted to achieve $\mu_k$ fantasy points from our predictive model.

\subsection{Feasibility Constraints}

We begin by formulating the basic feasibility constraints of a lineup.  Daily fantasy sports contests all impose a budget $B$ constraint on a lineup.  For DraftKings, this budget is $B=50,000$ fantasy dollars.   \text{As mentioned earlier, } there is also a constraint on the number of players of each type of position. {For notation,} we let the cost of player $j$ be $c_j$. There are $N_T$ NHL teams playing in a contest, and we denote the set of players on team $l$ by $T_l$. {In addition}, we denote the set of players who are centers, wingers, defenseman, and goalies by $C$, $W$, $D$, and $G$, respectively.    The DraftKings position constraints for the $i$th lineup are given by
\begin{equation}\label{eq:position_constraint}
\begin{aligned}
& & & 2\leq\sum_{j \in C} x_{ij} \leq 3, &\\
& & & 3\leq\sum_{j \in W} x_{ij} \leq 4, &\\
& & & 2\leq\sum_{j\in D}x_{ij} \leq 3, &\\
& & & \sum_{j\in G}x_{ij} = 1, &
\end{aligned}
\end{equation}
DraftKings also requires each lineup to have players that come from at least three different NHL teams.
The team constraints for lineup $i$ are
\begin{equation}\label{eq:team_constraint}
\begin{aligned}
& & & t_{il} \leq \sum\limits_{k \in T_l} x_{ik}, & \hunter{$l = 1, \ldots, N_T$} \\
&&& \sum\limits_{l=1}^{N_T} t_{il} \geq 3, & \\
&&& t_{il} \in \left\{0, \; 1 \right\},  & \hunter{$l = 1, \ldots, N_T$}.
\end{aligned}
\end{equation}

We put these constraints together to obtain the feasibility constraints for a lineup:
\begin{equation}\label{eq:feasible}
\begin{aligned}
&&&\sum_{j=1}^{p}c_jx_{ij} \leq B, ~~~\text{(budget constraint)}\\
& & & \sum_{k=1}^{p}x_{ij} = 9, ~~~\text{(lineup size constraint)}\\
&&& \text{Equations \eqref{eq:position_constraint},}~~~\text{(position constraint)}\\
&&& \text{Equations \eqref{eq:team_constraint},}~~~\text{(team constraint)}\\
&&& x_{ij} \in \curly{0,1}, ~~~ \hunter{$j = 1, \ldots, p$}.
\end{aligned}
\end{equation}

\subsection{Stacking Lineups Using Structural Correlations}\label{sec:stacking}
We now present the integer programming constraints on the lineup variance.  The constraints we use are simpler
than the general form in equation \eqref{eq:ip_general}.  Rather than force the variance of the lineup to
be larger than a lower bound, we constrain the structure of the lineup to ensure that its variance will be large.  We assume equal variance of all players{,} and covariances that are much smaller in magnitude than the variances.  This allows us to
approximate the variance of a lineup using only the number of pairs of positively  and negatively correlated players.  The constraints we present next, which we refer to as different types of stacking, exclude any negatively correlated players and try to have a large number of positively correlated players.

{Goalie Stacking.}  As discussed in Section \ref{sec:correlations}, goalies are negatively correlated with the skaters that they are opposing. Therefore, to increase the variance of a lineup, we would like to avoid having this negative correlation by making sure {none of the} lineup's skaters are opposing its goalie.  We will refer to this constraint as \emph{goalie stacking}. To model this constraint in our integer programming formulation, we denote the set of skaters who are opposing goalie $k$ by $O_k$.  For lineup $i$, the goalie constraint is then given by:
\hunter{
\begin{equation}\label{eq:no_goalie}
\begin{aligned}
& & &   \sum_{l\in O_j} x_{il} \leq 6(1-x_{ij}), & \forall j \in G.
\end{aligned}
\end{equation}
}
This constraint will force the goalie variable $x_{ik}$ to be zero if the lineup has any skater opposing goalie $k$.

{Line Stacking.}  Additionally, in Section \ref{sec:correlations} we found that players on the same line are positively correlated. Therefore, to increase the lineup variance, we would like to impose additional constraints on our formulation that will create lineups with players on the same line. We will refer to this as \emph{line stacking}.
Each entry consists of eight skaters, {six being offensive (wingers and centers), and a line consists of three offensive skaters.  The position constraints allow a lineup to include a maximum of two lines.}  To include as much positive correlation as possible amongst the players, we can make a lineup have one complete line of three players, and {one partial line with at least two players.  We do this by requiring a lineup have two partial lines and one complete line.  Given the position constraints, this will force one of the partial lines to be the same as the complete line.  }
To formulate the line stacking constraints, we introduce some notation. Let $N_L$ denote the total number of lines, and let $L_l$ denote the set of players on line $l$. With this notation we can formulate the constraint of having at least one complete line in  lineup $i$ as
\begin{equation}\label{eq:complete_line}
\begin{aligned}
& & 3v_{il} &\leq \sum\limits_{j \in L_l} x_{ij}, &\hunter{$l=1, \ldots, N_L$} &  \\
&& \sum\limits_{l =1}^{N_L} v_{il} &\geq 1 & \\
&& v_{il} &\in \left\{0,1 \right\},  & \hunter{$l=1, \ldots, N_L$}.
\end{aligned}
\end{equation}
Under these constraints $v_{il}$ is one if lineup $i$ has all three players from line $l$, and zero otherwise, and we require at least one of the $v_{il}$ to be one.
To formulate the constraint of having at least two players from two distinct lines, we use the following constraints:
\begin{equation}\label{eq:two_thirds_line}
\begin{aligned}
& &  2w_{il} &\leq \sum\limits_{k \in L_l} x_{ik}, & \hunter{$l=1, \ldots, N_L$}  \\
&& \sum\limits_{l =1}^{N_L} w_{il} &\geq 2 & \\
&& w_{il} &\in \left\{0,1 \right\}, & \hunter{$l=1, \ldots, N_L$}.
\end{aligned}
\end{equation}
Similar to the complete line constraints, this constraint sets $w_{il}$ to one if at least two out of three players of line $l$ are in  lineup $i$, and there must be at least two $w_{il}$ equal to one.

{Defensemen Stacking.}  Lastly, we have found that defensemen that are on the first power play line score substantially higher than defensemen that are not.  {While including the stronger defenseman may not affect the average points of the lineup, it could affect its actual points. If we force the lineup to have three teams, then the teammates of this defenseman will likely be included in the lineup.  This produces additional positive correlation within the lineup, which is desirable.  Because the defenseman is from the first line, when his team does well, we would expect the lineup to achieve more points than if we had chosen a weaker defenseman. }    We will refer to this as \emph{defensemen stacking}.
To model this constraint, let $P_D$ denote the set of defensemen that are on any team's first power play line. Then, to use only these defensemen in lineup $i$ we add the following constraints to our integer programming formulation:
\begin{equation}\label{eq:PP}
x_{ij} = 0 \quad \forall j \in D\setminus P_D.
\end{equation}
This constraint makes sure that any defenseman that is chosen for the lineup is also in the first power play line of some team.

There are many other constraints that could be added to the model, such as forcing a lineup to have every player on a team's first power play, imposing constraints to have seven players from one game, etc.  We did not investigate all the possible types of stacking, but they can all be naturally incorporated into our integer programming formulation.  In this work we focus on the constraints presented because they capture the major structural correlations in hockey.

\subsection{Overlap Constraints}\label{sec:overlap_constraints}
Our integer programming {formulation} in Section \ref{sec:greedyIP} had a {upper} bound on the covariance of lineup $i$ with the previous $i-1$ lineups.
We assume that the player variance is much larger in magnitude than the player {covariances}.  Using this assumption,
we approximate the correlation of lineups $i$ and $j$ as simply the overlap of the two lineups.  We can then
add constraints {to} {upper} bound this overlap with all previous $i-1$ lineups.  We will refer to these as  overlap constraints.  We let $x^*_{lj}$ be the optimal value of the variable for player $j$ in lineup $l$.  We define the overlap constraints for lineup $i$ as

\begin{equation}
\sum_{j=1}^{p}x^*_{lj}x_{ij} \leq \gamma, \quad  l = 1, \ldots, i-1.\label{eq:overlap}\\
\end{equation}
These constraints make sure that lineup $i$ will have no more than $\gamma$ players in common with any previously constructed lineup.
We refer to $\gamma$ as the maximum lineup overlap.  Decreasing $\gamma$ forces the lineups to be more diverse, whereas for larger $\gamma$
the lineups \text{may} share more players.  We will see in Section \ref{sec:performance} how the choice of $\gamma$
is impacted by the number of NHL games being played on a given night.

\subsection{Complete Sequential Integer Programming Approach}
We now present the {complete sequential integer programming approach} for constructing $k$ lineups.  We obtain lineup $i =1, \ldots, k$  by
solving
\begin{equation}\label{eq:ip_hockey}
\begin{aligned}
& \underset{x}{\text{maximize}} & & \sum_{j=1}^{p}\mu_jx_{ij} &\\
& \text{subject to} & &\text{Equation \eqref{eq:feasible}} ~~~\text{(feasibility constraints)}\\
&  & &\text{Equation \eqref{eq:overlap}} ~~~\text{(overlap constraints)}\\
&  & &\text{Any subset of equations \eqref{eq:no_goalie}}, \eqref{eq:complete_line},\eqref{eq:two_thirds_line}, \eqref{eq:PP}~~~\text{(stacking constraints)}.\\
\end{aligned}
\end{equation}
By solving this integer programming formulation iteratively $k$ times, each time updating the overlap constraints with the previously found solutions,
one obtains the $k$ greedy lineups.  This formulation allows flexibility in the choice of stacking constraints,
prediction model, maximum lineup overlap, and number of linueps constructed.  We will see next how to
select the best settings using data from real fantasy hockey contests.

\section{Performance}\label{sec:performance}
We now evaluate the performance of our {approach} on real daily fantasy sports hockey contests.
To obtain data on the performance of the population lineups in several actual {DraftKings} hockey contests
we {need} to enter the contests.  As entrants, we are able to download the lineups and points
of every lineup in the contest.  In total we have data for 38 different top-heavy contests {between October 21, 2015 and December 31, 2015,}  with {each containing}  several thousand entrants.    We  evaluate how much profit our lineups would have made in each contest competing against the actual population lineups.  We investigate a variety of settings, including the number of lineups, type of stacking,  prediction model, and maximum overlap.

We combine the different constraints from Section \ref{sec:stacking} to create six different types of stacking{,} which are summarized  in Table \ref{table:stacking}.   The first type is no stacking, which only uses the basic constraints and none of the stacking constraints.   All the other types of stacking use goalie stacking (equation \eqref{eq:no_goalie}) as this is a natural negative correlation we wish to avoid.  {In addition}, all the other types of stacking require at least one complete line (equation \eqref{eq:complete_line}), and at least two partial lines (equation \eqref{eq:two_thirds_line}). Only two defensemen are allowed in Type 2, the logic being that the utility player should be a {forward} that will score more points, as was seen in Figure \ref{fig:figure_points_line_error}.  Types 3 and 4 use defensemen stacking (equation \eqref{eq:PP}).  Types 4 and 5 require exactly three different teams to be represented on a lineup.  This constraint increases the lineup variance by having more players on the same team in the lineup.

\begin{table}
\begin{center}
\begin{tabular}{| c | c | c | c | c | c |}
\hline No stacking & Type 1 & Type 2 & Type 3 & Type 4 & Type 5 \\ \hline
$\begin{aligned}[t] \text{Equation \eqref{eq:feasible}} \end{aligned}$&
$\begin{aligned}[t]
\text{Equation \eqref{eq:feasible}}\\
\text{Equation \eqref{eq:no_goalie}} \\
\text{Equation \eqref{eq:complete_line}} \\
\text{Equation \eqref{eq:two_thirds_line}}
\end{aligned} $ &
$\begin{aligned}[t]
\text{Equation \eqref{eq:feasible}} \\
\text{Equation \eqref{eq:no_goalie}} \\
\text{Equation \eqref{eq:complete_line}} \\
\text{Equation \eqref{eq:two_thirds_line}} \\
\sum_{j\in D}x_{ij} = 2
\end{aligned}$ &
$\begin{aligned}[t]
\text{Equation \eqref{eq:feasible}}\\
 \text{Equations \eqref{eq:no_goalie}} \\
 \text{Equations \eqref{eq:complete_line}} \\
 \text{Equations \eqref{eq:two_thirds_line}} \\
 \text{Equations \eqref{eq:PP}}
 \end{aligned}$ &
$\begin{aligned}[t]
\text{Equation \eqref{eq:feasible}}\\
\text{Equations \eqref{eq:no_goalie}} \\
\text{Equations \eqref{eq:complete_line}} \\
\text{Equations \eqref{eq:two_thirds_line}} \\
\text{Equations \eqref{eq:PP}} \\
\sum\limits_{k=1}^{N_T} t_{ik} = 3\\
\sum_{k \in T_l} x_{ik} \leq 6t_{il},\\ \; \;  l=1,\ldots, N_T \\
\end{aligned} $&
$\begin{aligned}[t]
\text{Equation \eqref{eq:feasible}}\\
\text{Equations \eqref{eq:no_goalie}} \\
\text{Equations \eqref{eq:complete_line}} \\
\text{Equations \eqref{eq:two_thirds_line}} \\
\sum\limits_{k=1}^{N_T} t_{ik} = 3\\
\sum_{k \in T_l} x_{ik} \leq 6t_{il},\\ \; \;  l=1,\ldots, N_T \\
\end{aligned} $\\ \hline
\end{tabular}
\end{center}
\caption{Summary of the  constraints for different types of stacking in our integer programming formulation for constructing hockey lineups.}\label{table:stacking}
\end{table}
\subsection{Impact of Stacking and Number of Lineups}\label{sec:number}
\begin{figure}\centering
    \includegraphics[scale=.3]{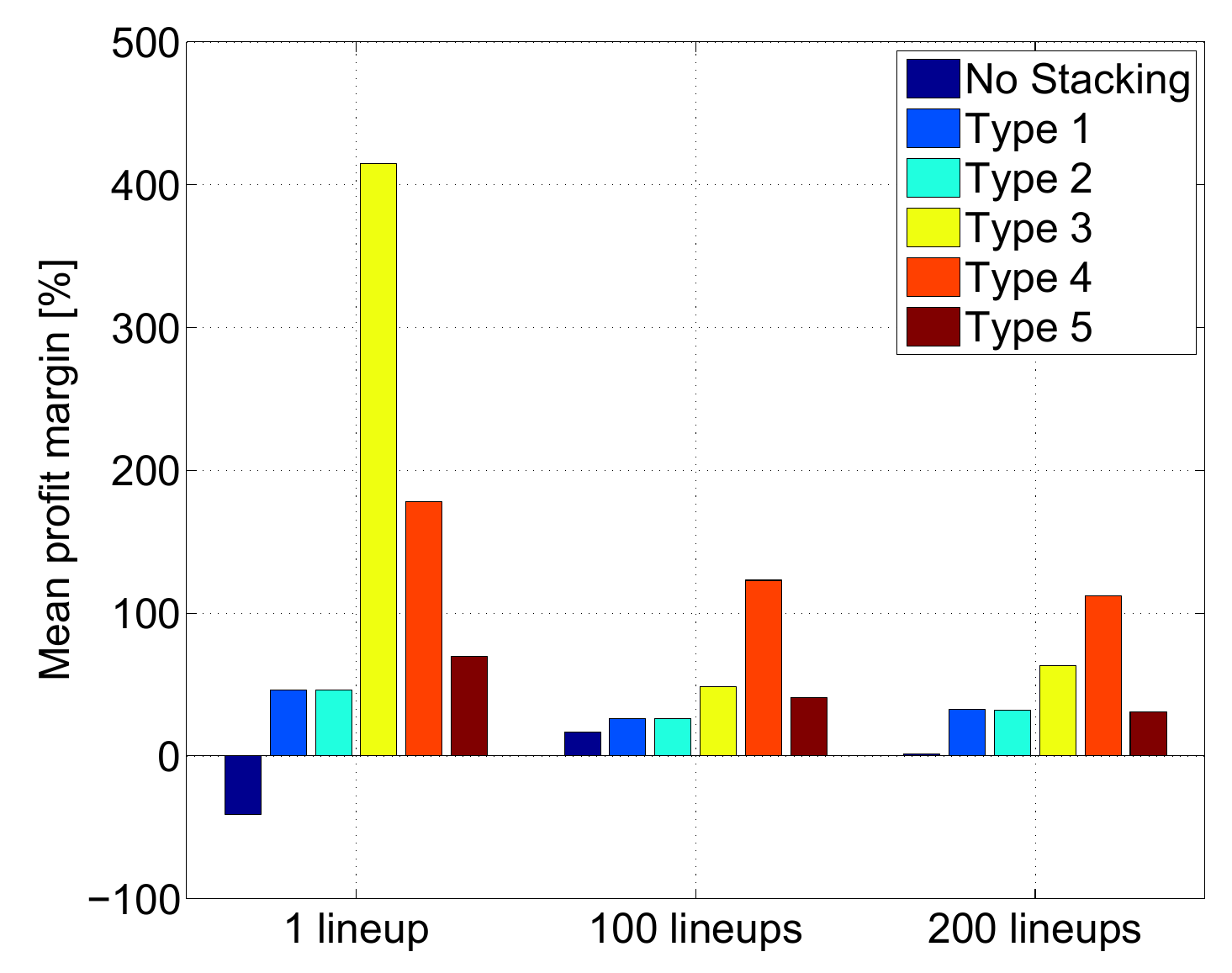}
    \caption{Plot of the mean profit margin in DraftKings hockey contests versus the number of integer programming lineups.  The plots are for  different types of stacking constraints.  The maximum lineup overlap allowed is seven.   }
    \label{fig:stacking}
\end{figure}

\begin{figure}\centering
    \includegraphics[scale=.3]{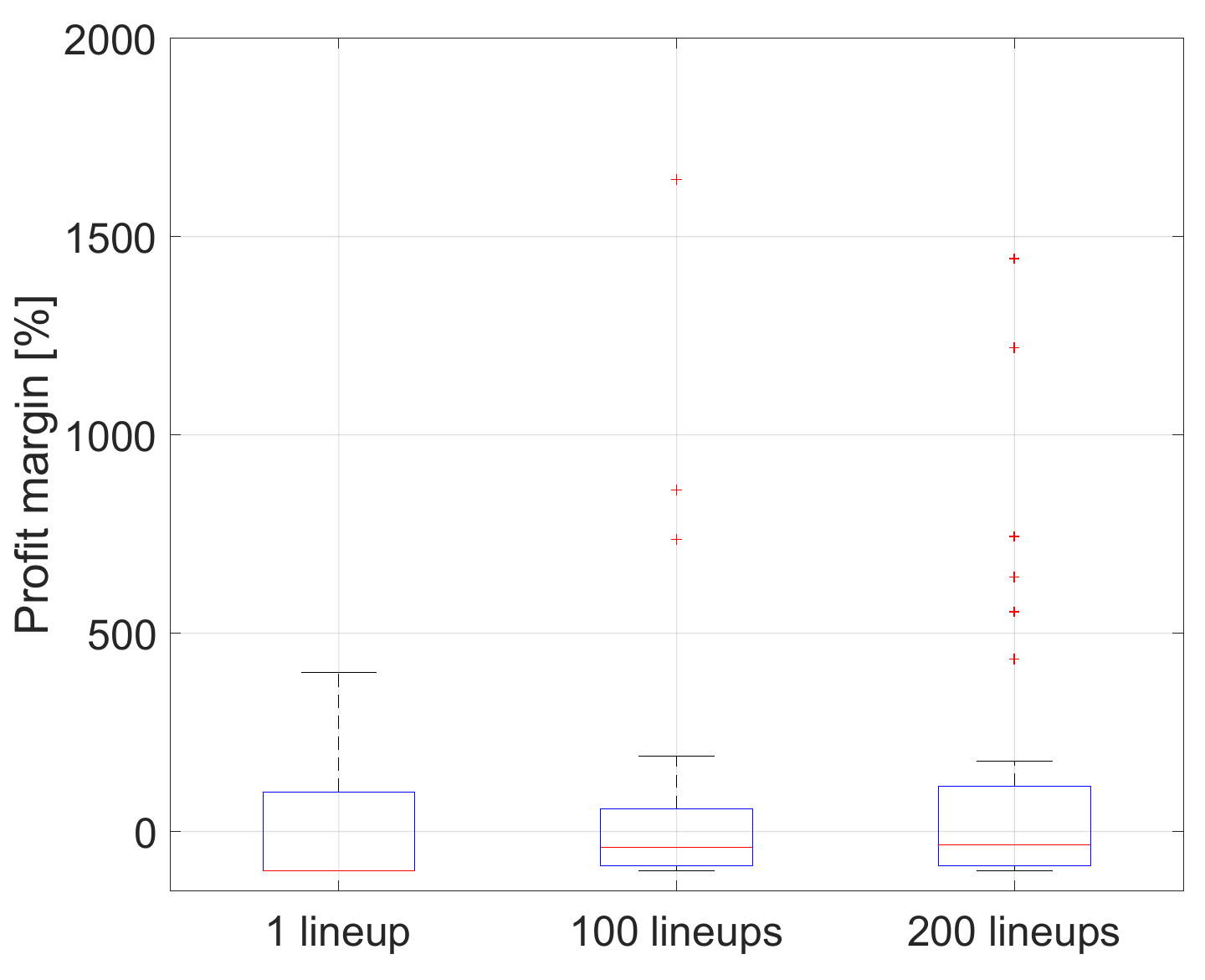}
    \caption{Boxplot of the profit margin in DraftKings hockey contests versus the number of integer programming lineups.  The lineups use Type 4 stacking and a maximum lineup overlap of seven.   }
    \label{fig:lineups}
\end{figure}

We begin by investigating the performance of each type of stacking versus the number of lineups entered.
We create different numbers of lineups for each type of stacking with a maximum lineup overlap of seven and calculate the mean profit margin across all contests. {In calculating the mean profit margin, we must adjust for the fact that when we enter more lineups, there will be some entries that cannot be in the contest pool now because of DraftKings lineup entry limits. To adjust for this, we randomly sample lineups to delete from the historical data as we increase the number of lineups we enter, and we report our mean profit margin across one-thousand trials of this random process.} We plot the results in Figure \ref{fig:stacking}.  We observe that the lowest profit margin is achieved by no stacking.  This shows the importance of stacking lineups to increase lineup variance.  Type 4 stacking achieves the highest profit margin for 100 and 200 lineups.  For one lineup, Type 3 stacking has a higher mean profit margin due to a single contest where the profit margin is quite large.  Excluding this single outlier, Type 4 stacking has the highest mean profit margin for one lineup.  Recall that Type 4 stacking includes goalie stacking, line stacking (complete and partial), defensemen stacking, and at most three different teams.  It has more stacking than all other types, which we expect to give its lineups more variance.  We see from the data that this maximizing variance strategy is effective for winning top-heavy contests.

 It seems from Figure \ref{fig:stacking} that the mean profit margin is largest for one lineup.  However, this is a deceiving statistic because the mean is dictated essentially by a single outlier for these top-heavy contests.  A more accurate picture is provided by the boxplot in Figure \ref{fig:lineups}.  Here we show the profit margin for Type 4 stacking with a maximum lineup overlap of seven.  We see that with one lineup, the median profit margin is -100\%.  However, with more lineups, the median profit margin becomes larger.  In addition, the 75th percentile is larger for  200 lineups than for 100 lineups.  {Moreover, for 100 and 200 lineups it can be seen there are many outliers that produce much of the profit, whereas with one lineup there are no outliers.}  This shows that with more lineups the distribution of the profit margin is shifted upward and given a slightly fatter tail.  Therefore, to win more money consistently, it is advantageous to use more lineups {(at least up to the limit imposed by real-life contests)}.

\begin{figure}\centering
    \includegraphics[scale=.4]{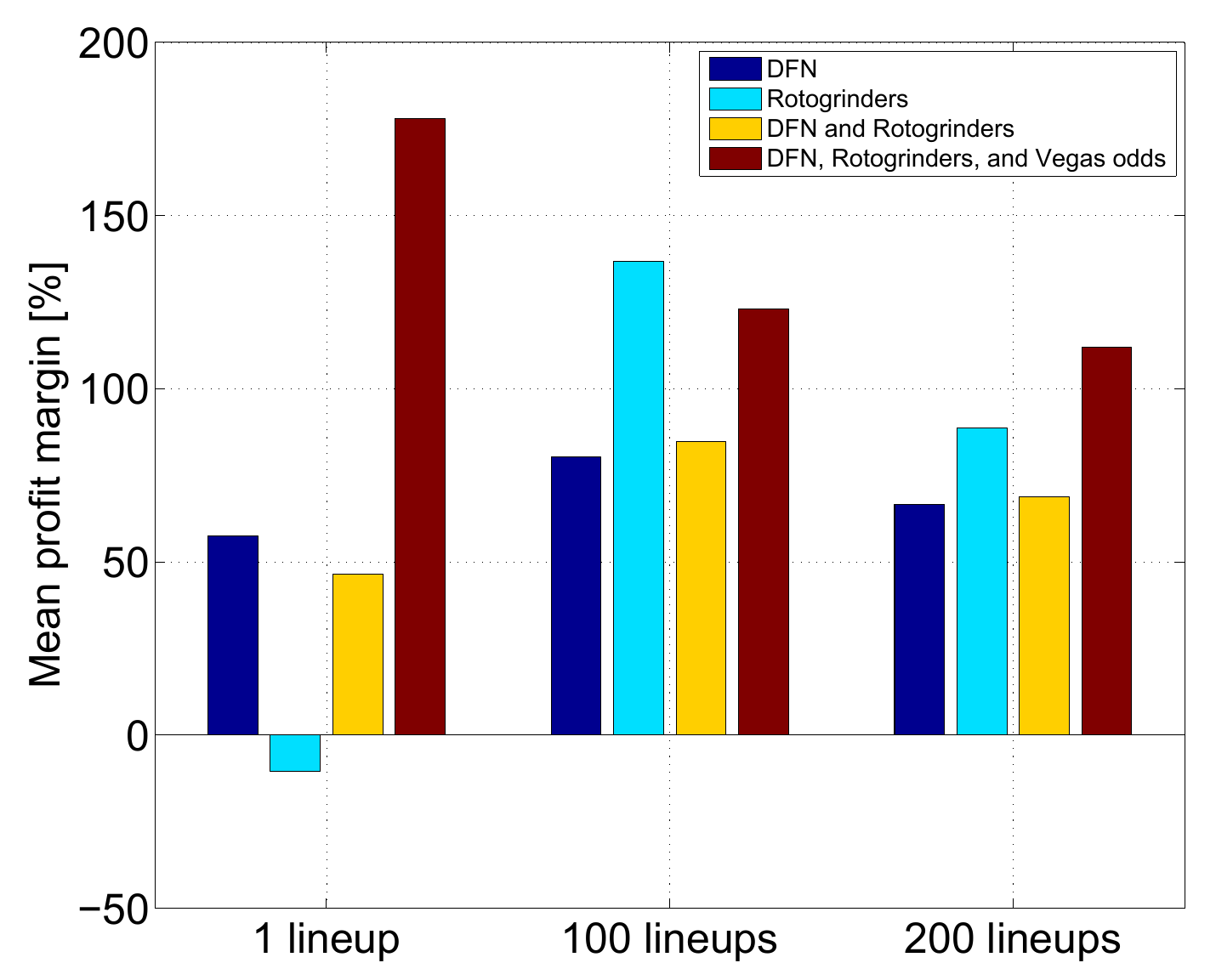}
    \caption{Plot of the mean profit margin of our integer programming approach in DraftKings hockey contests with different prediction models.  The models considered are Daily Fantasy Nerd predictions (DFN), Rotogrinders predictions, DFN and Rotogrinders predictions, and DFN and  Rotogrinders predictions with Vegas odds.  The maximum lineup overlap allowed is seven and constraint type four is used for the integer programming lineups.   }
    \label{fig:prediction}
\end{figure}

\subsection{Impact of Prediction Models}
{We next investigate the impact of the different prediction models we developed in Section \ref{sec:prediction} on our performance.}  We consider Type 4 stacking, a maximum lineup overlap of seven, and 200 lineups.  The prediction models we consider use Rotogrinders predictions, Daily Fantasy Nerd predictions, both sites' predictions, and both sites plus the win probability.  We show in Figure \ref{fig:prediction} the mean profit margin of different prediction models for different numbers of integer programming lineups.  We see here that the model with all three features does the best or near the best for each number of lineups.  The reason here may be that by including the win probability, the integer programming lineups are more biased towards goalies expected to win the games, resulting in a better chance of obtaining the game win bonus of three points.  {However, we saw in Section \ref{sec:prediction} that the regression models had poor performance in terms of predictive accuracy.  Therefore, it is not clear from our analysis which model has the clear advantage.   }

\subsection{Impact of Maximum Lineup Overlap}
\begin{figure}\centering
    \includegraphics[scale=.5]{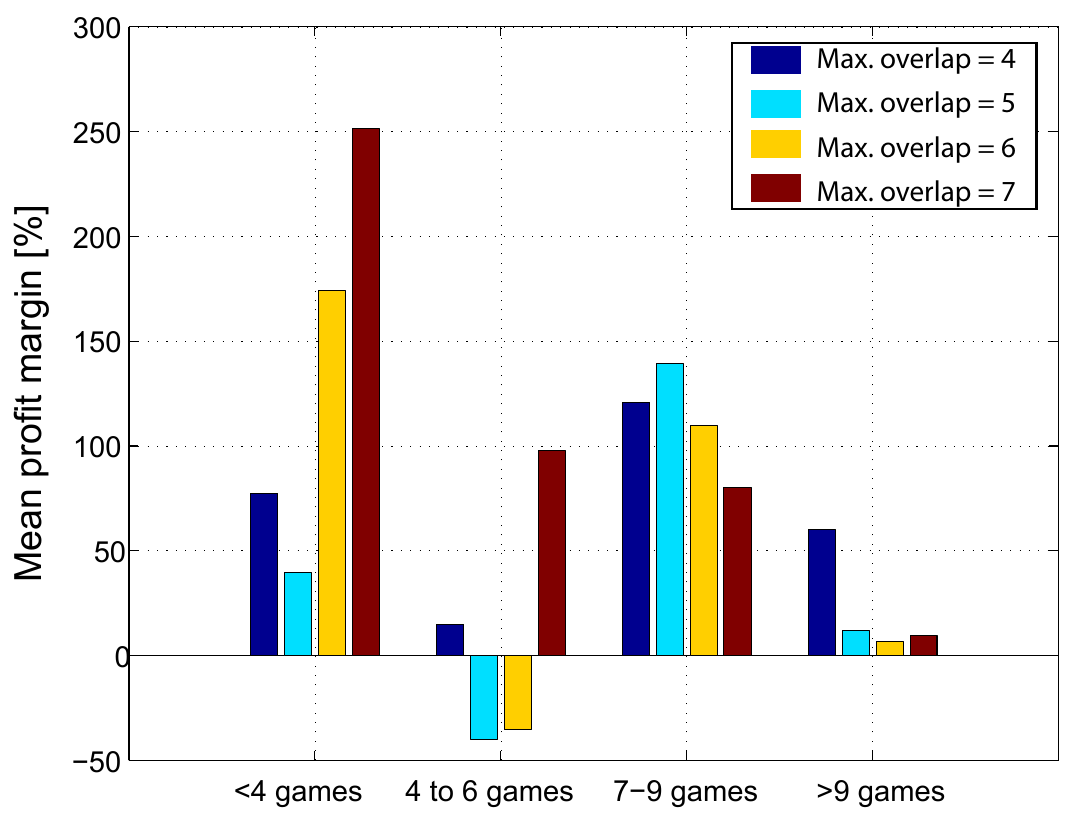}
    \caption{Plot of the mean profit margin 200 integer programming lineups with Type 4 stacking in DraftKings hockey contests versus maximum lineup overlap for different number of NHL hockey games being played.   }
    \label{fig:profit_games}
\end{figure}
One parameter we can adjust in our model is the maximum lineup overlap.  By decreasing the allowed overlap, we force the lineups to be more diverse.  We find that the degree of diversity we want depends upon how large the space of lineups is.  The size of the lineup space on a given night is larger if there are more NHL games being played.  We plot the mean profit margin for 200 integer programming lineups as a function of the number of games played in a night for different values of the maximum lineup overlap in Figure \ref{fig:profit_games}.  These lineups were created using Type 4 stacking.  For nights with {fewer} than four games, a maximum lineup overlap of seven does best.  For a larger number of games, it seems that decreasing the maximum lineup overlap improves performance.  For instance, for nights with more than nine games, an overlap of four does best.  The reason for this is that when there are few games being played, the lineup space is smaller and probably has fewer good lineups.  By reducing the maximum lineup overlap, we end up not selecting these good lineups and instead choose many poor lineups.  However, on nights with many games, there is a huge lineup space that has several different good lineups. A smaller maximum lineup overlap causes the integer program to choose several of these good lineups, increasing the chance that one of them will achieve a high rank.  For maximum lineup overlaps smaller than four we found that it becomes difficult to obtain 200 feasible lineups.  Maximum  lineup overlaps larger than seven produce lineups that are too similar and do not provide enough diversity.  Therefore, the maximum lineup overlap should be tuned between four and seven depending on how many games are being played on a given night.

\subsection{Lineup Creation Order Versus Lineup Performance Rank}
\begin{figure}\centering
	\includegraphics[scale=.3]{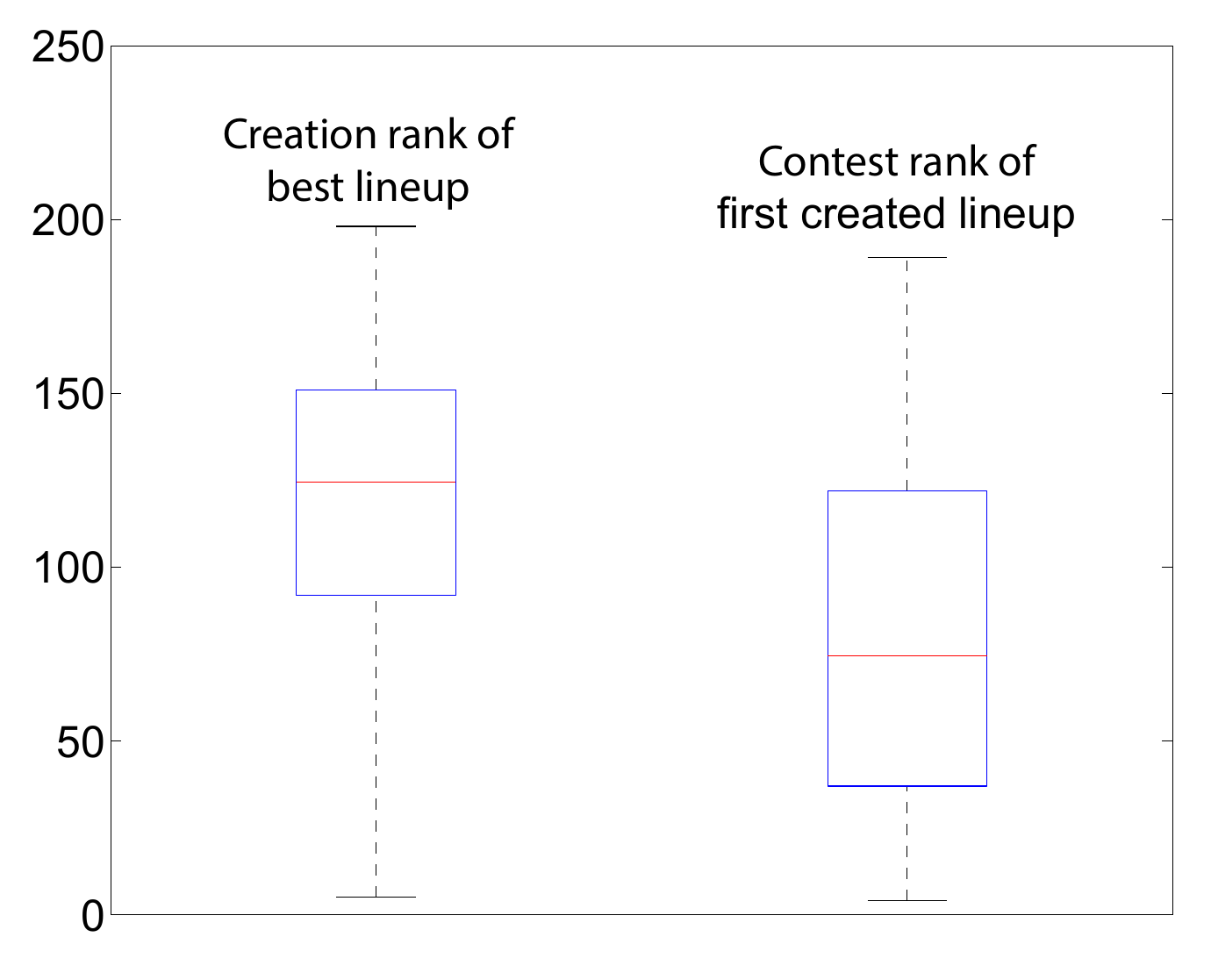}
	\caption{Plot of the performance rank of the first created lineup and the creation rank of the best performing lineup.  There were 200 integer programming lineups created with Type 4 stacking and a maximum overlap of seven.   }
	\label{fig:rank}
\end{figure}

{In constructing our lineups, each one is constructed sequentially to create multiple lineups.}  One question to ask is do the lineups {that} are created earlier perform better?  To investigate this, we set the integer programming approach to create 200 {lineups} with Type 4 stacking and a maximum overlap of seven.  We then evaluate their performance rank relative to each other in the contests.  This allows us to look at several statistics relating the creation rank and the performance rank of the lineups.  We first investigate the Spearman rank correlation coefficient between these two rankings.  We find that the average value of the correlation coefficient across the different contests is 0.09 with a standard deviation of 0.1.  This suggests that there is little correlation between the creation rank and performance rank of the lineups.  To dive deeper into the analysis, we plot in Figure \ref{fig:rank} two different sets of data.  First, we show a boxplot of the creation rank of the best performing lineup. Second, we show the performance rank of the first created lineup.  We see that the median performance rank of the first created lineup is 74.5.  This shows that the first created lineup has a slightly, but not substantially improved performance.  For the best performing lineup, the median creation rank is 124.5.  Therefore, while the first created lineup is slightly better, the winning lineup is generally one of the later created lineups.  This supports our finding of a low correlation between the creation and performance rank of the lineups.

\subsection{Mean and Standard Deviation of Lineup Points Versus Profit Margin}
\begin{figure}\centering
	\includegraphics[scale=.5]{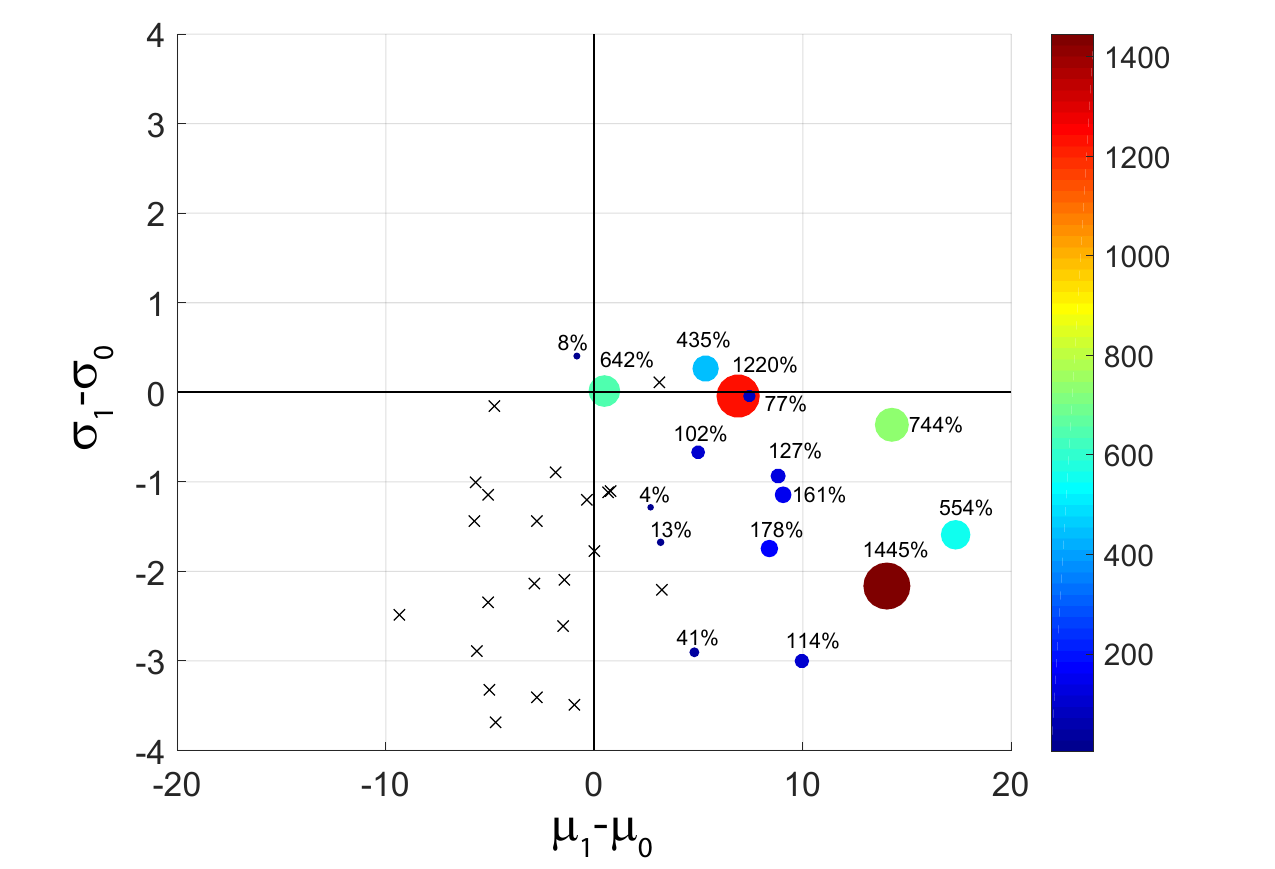}
	\caption{Plot of profit margin versus $\mu_1-\mu_0$ and $\sigma_1-\sigma_0$ for DraftKings hockey contests  with 200 integer programming lineups created using Type 4 stacking and a maximum overlap of seven. The profit margin for each contest is indicated through the color and size of the markers and also listed next to the marker.  The \hunter{$\times$} markers indicate contests where the profit margin is not positive.}
	\label{fig:difference}
\end{figure}

One way to characterize the lineups produced by our approach is by the mean and standard deviation of their points.  We can compare these with the mean and standard deviation of the points of all lineups in a contest.  Let us define the mean and standard deviation of our optimized lineups as $\mu_1$ and $\sigma_1$ and of the population lineups as $\mu_0$ and $\sigma_0$.  We will look at the difference in these parameters for our lineups and the population lineups for each contest.  We use 200 integer programming lineups with Type 4 stacking and a maximum overlap of seven and plot the profit margin versus the difference in the mean and standard deviation in Figure \ref{fig:difference}.  One can see that there is a substantial variation in mean difference, which ranges from -10 to 20 points.  When it is below zero, no profit is made, but when it is above zero, a substantial profit is made.  This is a result of the fact that the lineups must satisfy the stacking constraints {that} increase their individual variance.  The result is that when the actual points are realized, we are either far above or far below the population mean.  The standard deviation of the integer programming lineups is generally below the population standard deviation.  This is because we are choosing lineups that have some correlation because they are designed to maximize points, even though the individual lineups are designed to have a high variance.     From Figure \ref{fig:difference} one can see that we are equally likely to be either above or below the population mean.  However, because the contest payoff structure is so asymmetric, with a maximum loss of 100\%, but a maximum gain {that} is at least an order of magnitude larger, our {lineups} end up being profitable.

\subsection{Implementation and Runtime}

All formulations were constructed using the  JuMP algebraic modeling language  \citep{DunningHuchetteLubin2017,LubinDunningIJOC}, which is written in the Julia programming language \citep{Bezanson:2012}. This allowed us to test the runtime of our approach for various integer programming solvers. {The complete software used to construct the integer programming lineups is available at {an open-source GitHub webpage \citep{ref:github_hockey}}. }

\begin{figure}\centering
	\includegraphics[scale=.35]{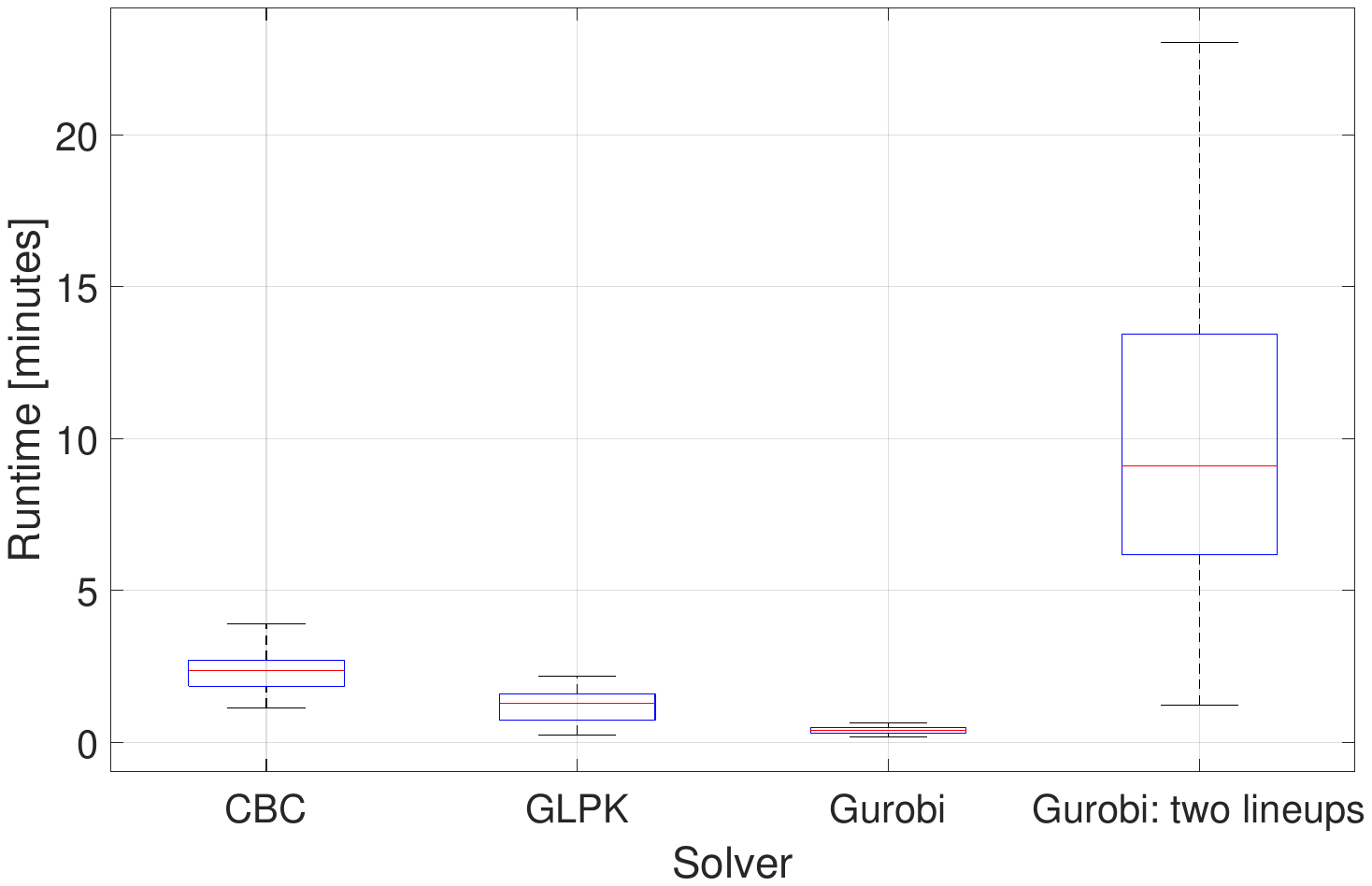}
	\caption{The runtime of our algorithm for generating 100 lineups using Type 4 constraints with an overlap parameter of 6 for 38 different hockey contests with different integer programming software.  The box labeled ``Gurobi: two lineups'' is the runtime when solving two lineups at a time using Gurobi.  The other boxes are for the sequential approach where we solve for one lineup at a time.  }
	\label{fig:runtime}
\end{figure}

In practice, we must wait for information about {the} goalies {that} are {playing} to be posted before we construct our lineups.  Goalies do not play every night, so if we do not wait for this information, we may put goalies in lineups who are not playing  and receive zero points for them.  This goalie information is generally posted on public websites about 30 minutes before the games begin.  We must be able to {find a solution} in this time to be able to enter the contest. {The first three plots} in Figure \ref{fig:runtime} {show} the time needed to solve the {algorithm} {presented in Section \ref{sec:algorithm} with Type 4 constraints and a maximum lineup overlap of six} for 100 lineups in each of our contests using different integer programming solvers. We consider the free solvers CBC \citep{cbc} and GLPK \citep{glpk} and the commercial solver Gurobi \citep{gurobi}. All computations were done using a Intel Core i5-4570 3.20GHz with 8GB of RAM.  We find that all solvers are able to solve {our algorithm for 100 lineups} in under four minutes, giving sufficient time to enter the lineups into the contest.

{While our simple formulations were more than adequate for the approach that allowed us to consistently win over several weeks, solve times could become an issue when more elaborate approaches are used. For instance, instead of solving for one lineup at a time like we propose, we could solve for two lineups at a time. To consider lineups $i$ and $i+1$ we simply combine the $i$-th formulation (on variables $\{x_{ij}\}_{j=1}^{p}$) with the $i+1$-th formulation (on variables $\{x_{(i+1)j}\}_{j=1}^{p}$) by adding a variant of the overlap constraint \eqref{eq:overlap} which restricts the overlap between the two lineups and is given by}
\begin{equation}
\hunter{$\sum_{j=1}^{p}x_{(i+1)j}x_{ij} \leq \gamma$}\label{eq:overlaptwo}.\\
\end{equation}
{Unlike the original overlap constraint \eqref{eq:overlap}, constraint \eqref{eq:overlaptwo} is a quadratic constraint similar to the first set of constraints of  formulation \eqref{eq:ip_general}. As described in Appendix~\ref{formulation_appendix} these constraints can be linearized through standard techniques. However, the resulting formulation can be much harder to solve. In the fourth box in Figure \ref{fig:runtime}, we show the amount of time it takes to solve for 100 lineups when solving lineups two at a time using Gurobi. From this plot, it is clear that solving for lineups two at a time leads to a substantial increase in runtime. We also considered solving for two lineups at a time with the free solvers CBC and GLPK, but we found that it frequently takes longer than two hours to determine the 100 lineups. Fortunately, over 50 years of integer programming research provide plenty of opportunities to improve solve times of this and other extensions \citep{Juenger2010}. In particular, we should note that our specific implementation of the Goalie Stacking constraint and the full Line Stacking  constraints (i.e. \eqref{eq:no_goalie} and \eqref{eq:complete_line}) are examples of what classical textbooks call \emph{weak formulations} and are often associated to  increased solve times. These formulations can be strengthened through standard techniques that are detailed in Appendix~\ref{formulation_appendix}. However, nowadays  such strengthening can lead to increased solve times, partly because of an increased number of constraints and partially because of the idiosyncrasies of modern solvers \citep[page 69]{conforti2014integer}. Indeed, for our instances this  strengthened version resulted in rather significant  solve-time increments for GLPK and no noticeable improvements for Gurobi and Cbc (See Appendix~\ref{formulation_appendix}). Nonetheless, more advanced formulation strengthening techniques are regularly used to increase the performance of modern solvers with very low implementation cost so they should not be automatically ignored \citep{Mixed-Integer-Linear-Programming-Formulation-Techniques}.     }

\subsection{Baseball}\label{sec:baseball}
Having demonstrated the success of our  approach in hockey contests, {a natural goal would be to consider how much our approach generalizes.}  To do this, we \text{apply} our  approach to top-heavy daily fantasy baseball contests in DraftKings. {For baseball we only include results from May and June of 2016, so we have less data for this sport than hockey.} {The} structure
of baseball contests is similar to hockey contests.  Therefore, {for brevity} we will present here {only} a brief summary of our baseball {approach} and its performance on real contest data.  In short, we have found that in our limited number of baseball contests, the method does indeed perform  well and is able to place in the top ten in contests with tens of thousands of entrants multiple times.

We begin with a brief overview of how baseball is played and how fantasy points are scored.  A baseball team consists of pitchers and hitters.  The goal is for the hitters to hit the ball pitched by the opposing pitcher and then run to four bases to score runs.  In fantasy sports contests a pitcher scores points for striking out batters, pitching for more innings in the game, and winning the game.  A hitter scores points by hitting the pitch, scoring a run by reaching home base, or having other hitters score a run as a result of his hit, also known as a run batted in or RBI.

  There is a great deal of similarity between the structure of hockey and baseball lineups, so we are able to almost directly use the hockey integer programming formulation with some slight modifications.   We do not go through the details of the baseball formulation here, but instead present the basic constraints.  First, our objective is to maximize the expected fantasy points of the lineups.  Here we simply use the projections from Daily Fantasy Nerd for the expected players' points.   Second, there are feasibility constraints related to the players' position, team, and salary.  A baseball lineup consists of ten players:  two pitchers and eight hitters (catcher, first baseman, second baseman, third baseman, shortstop, and three outfielders).  The lineup must have players from at least two different baseball games and the budget for a lineup is 50,000 fantasy dollars, just as in hockey.  Finally, a lineup cannot have more than five hitters from one team.

We have stacking and overlap constraints for baseball.  As in hockey, we want to stack the baseball lineups to increase their variance.  We do this by adding constraints {that} remove negative correlations and increase positive correlations within a lineup.   We use two different stacking types to do this.  First, similar to goalie stacking in hockey, we do not have a pitcher and any opposing players on the same lineup. Pitchers and opponents have negatively correlated points, just as in hockey with goalies and opposing skaters.
 Second, similar to line stacking in hockey, we use consecutive hitters on a lineup.  Consecutive hitters have positively correlated points because of the structure of baseball.  For instance, if the first three batters can get on base and the fourth batter hits a home run, then the first three batters get points for runs and the fourth batter gets points for his hit plus the {four RBI}.  In our formulation, we use five consecutive batters in each lineup, allowing for cyclic orders such as batters in positions $(8,9,1,2,3)$.  We choose five consecutive batters because this is the maximum allowed by DraftKings.
The overlap constraint has the same structure for both baseball and hockey except that we set the maximum overlap parameter $\gamma$ to six for baseball.
\begin{table}
\begin{center}
\begin{tabular}{ |c|c|c| }
 \hline
Date        & Best Rank & Total Number of Entries\\
\hline
5/25/2016 & 1    & 47,916\\ \hline
5/26/2016 & 3      & 38,333\\ \hline
5/27/2016 & 1,342 & 57,500\\ \hline
5/29/2016 & 7      & 38,333\\ \hline
5/30/2016 & 213  & 38,333\\ \hline
5/31/2016 & 146  & 47,916\\ \hline
 6/1/2016 & 400  & 47,916\\ \hline
 6/2/2016 & 17   & 46,000\\ \hline
 6/3/2016 & 376  & 53,666\\ \hline
 6/5/2016 & 2      & 38,333\\
 \hline
\end{tabular}
\caption{The date, our top-ranked lineup, and number on entries in DraftKings top-heavy baseball contests.}
\end{center}
\end{table}

Baseball contests allow 200 entries per person, but are typically larger than hockey contests, which makes them more difficult to win.  While top-heavy hockey contests never have more than 24,000 entrants,  top-heavy baseball contests always have more than 38,000 entrants.   Nonetheless, our approach has proven successful .  We show the rank achieved by our best lineup (out of 200) in historic DraftKings daily fantasy baseball competitions and the number of entrants in the competitions in Table 4.  As can be seen, our {approach} is able to place in the top ten in four competitions, and even comes in first place in one of the competitions.  This success in baseball is significant because we use the same general principles presented in Section \ref{sec:greedyIP} {that} were used for hockey.  This suggests that our approach to picking winners
is general and has applicability to a variety of domains.

%

\section{Conclusion}\label{sec:conclusion}
We have presented here {an approach} for constructing entries
from a set of constrained resources, which have a large probability of winning
 top-heavy contests.
Our approach is developed from insights gained
from analyzing properties of the probability of winning, which is many times a {difficult function to evaluate}.
We have shown that our approach can consistently win top-heavy daily fantasy hockey and baseball contests.
While we {apply} our approach to one application domain, it is general and can be applied
in many other settings.  At a high level, our approach can be summarized as follows.
\begin{enumerate}
	\item Build simple prediction models for the resources' means and understand their pairwise correlations.
	\item Formulate an integer program that constructs entries with maximal mean while
	lower bounding their variance and upper bounding their correlation with previously constructed entries.
	\item Solve the integer programs to sequentially generate entries in a greedy manner.
\end{enumerate}
The success we were able to achieve for both hockey and baseball given the relative simplicity of our
approach shows its power.  We do not use the actual value of any correlations, just a basic understanding of their sign.
Our predictions can even be noisy.  It is the manner in which the entries are constructed {that}
allows us to pick winners despite noise in the prediction models.

Our success in using integer programming in daily fantasy sports has created greater
awareness for analytics and optimization.  Daily fantasy sports is a great application
to generate interest in these topics because it is a popular hobby and has a structure that
is ideal {for} integer programming.  Our initial results and code have already
exposed a large audience to integer programming and demonstrated how easy it is to use.
We feel that going forward, daily fantasy sports can be a great educational tool
for teaching analytics, optimization, and in particular integer programming.


\section{Proofs}\label{sec:proofs}
	\subsection{Proof of Theorem \ref{thm:nphard}}
	We show that maximizing $U(\mathcal S)$ is NP-hard by reducing it to the maximum coverage problem.
	In the maximum coverage problem one is given a set $\mathcal U$ of $n$ elements and a collection $\mathcal E =\curly{E_i}_{i=1}^N$ of
	$N$ subsets of $\mathcal U$ such that $\bigcup_{E\in\mathcal E}E=\mathcal U$.   The goal is to select $k$ sets from $\mathcal E$ such that their union has maximum cardinality.  This is known to
	be an NP-hard optimization problem.  To show that this is an instance of maximizing $U(\mathcal S)$ we assume
	that the sample space $\Omega$ is countable and finite with $R$ elements.  We also assume that each element $\omega\in\Omega$
	has equal probability, i.e. $\mathbf P(\omega)=R^{-1}$. Let $\mathcal F$ be the $\sigma$-algebra of $\Omega$.   For any set $\mathcal S\in\mathcal F$, we can write $U(\mathcal S) = R^{-1}|\bigcup_{\omega\in\mathcal S}\omega|$.
	Then we have
	\begin{align*}
	\max_{\mathcal S\subseteq \mathcal E, |\mathcal S|=k} U(\mathcal S) & = \max_{\mathcal S\subseteq \mathcal E, |\mathcal S|=k}R^{-1}\left|\bigcup_{\omega\in\mathcal S}\omega\right|\\
	&=\max_{\mathcal S\subseteq \mathcal E, |\mathcal S|=k}\left|\bigcup_{\omega\in\mathcal S}\omega\right|.
	\end{align*}
	Therefore, maximizing $U(\mathcal S)$ is equivalent to the maximum coverage problem.
	\subsection{Proof of Lemma \ref{thm:submodular}}
	The function $U(\mathcal S)$ is non-negative and non-decreasing because it is the probability of a set of events.
	We must show that it is also submodular.  A submodular function $f$ satisfies
	\begin{align}
	f\paranth{\mathcal S \bigcup v} - f\paranth{\mathcal S } &\geq f\paranth{\mathcal T \bigcup v}  - f\paranth{\mathcal T}
	\end{align}
	for all elements $v$ and pairs of sets $\mathcal S$ and $\mathcal T$ such that
	$\mathcal S\subseteq \mathcal T$.
	We show that the function $U(\mathcal S)$ is submodular as follows.  We let the $\sigma$-algebra of the probability space be $\mathcal F$.   Consider sets $\mathcal S,\mathcal T, v\in\mathcal F$ such that $\mathcal S\subseteq \mathcal T$.
	We can write $v=v_\mathcal S \bigcup v_\mathcal T\bigcup v_0$ where we define $v_\mathcal S = v\bigcap \mathcal S$, $v_\mathcal T = v\bigcap \mathcal T\bigcap \mathcal S^c$, and $v_0 = v\bigcap \mathcal T^c$.  Then we have
	\begin{align}
	U\paranth{\mathcal T\bigcup v}-U\paranth{\mathcal T}&= \mathbf P\paranth{v_0}\nonumber
	\end{align}
	and
	\begin{align}
	U\paranth{\mathcal S\bigcup v}-U\paranth{\mathcal S}&= \mathbf P\paranth{v_\mathcal T \bigcup v_0}\nonumber\\
	&\geq \mathbf P\paranth{v_0}\nonumber\\
	&\geq 	U\paranth{\mathcal T\bigcup v}-U\paranth{\mathcal T},\nonumber
	\end{align}
	thus showing that $U(\mathcal S)$ satisfies the submodularity condition.

\subsection{Proof of Theorem \ref{thm:U2}}
{
We begin by establishing a lower bound for $U(S)$.  For $k=1$ we have that $U(S) = p$ and for $k=2$ we have that
$U(S)= 2p-p^2$.  For $k\geq 3$ we establish a lower bound as follows.
Let use define $f(p)=1-(1-p)^k$, so $U(S) = f(p)$.  Then we form a second order Taylor series expansion for $f(p)$ about $p=0$,
\begin{align*}
    f(p) & = kp -\frac{1}{2}k(k-1)p^2 + R_3(p)
\end{align*}
where $R_3(p)$ is the Taylor series remainder.  For $k=1,2$ the remainder is zero.  For $k\geq 3$ it is given by
\begin{align*}
    R_3(p) = \frac{1}{4!}k(k-1)(k-2)(1-\xi)^{k-3}p^4
\end{align*}
for some $\xi\in[0,p]$.  Because $p<1$, we have that $R_3(p)>0$, which gives the lower bound
\begin{align*}
    U(S) \geq kp -\frac{1}{2}(kp)^2.
\end{align*}
This lower bound also holds for $k=1,2$.
}

{
The upper bound for $U(S)$ is obtained using Bernoulli's inequality, which states that $(1+x)^k\geq 1+kx$ for $x\geq -1$ and integer $k\geq 0$.  By applying this inequality we obtain
\[U(S) = 1-(1-p)^k\leq kp.\]
}

Next we establish lower and upper bounds on $U(S)-U_2(S)$.  To do this we rely upon the following lemma.
\begin{lemma}\label{lem:U2_pt}
For a set of entries $\mathcal S$, let $k=|\mathcal S|$.  Then
\begin{align}
	U_2(\mathcal S) & = \frac{1}{2}\sum_{l=1}^k l(3-l)\sum_{T\in \mathcal S_l}p'_T.
\end{align}
\end{lemma}
With this lemma we can rewrite $U_2(\mathcal S)$ as
\begin{align*}
	U_2(\mathcal S) & = U(\mathcal S) - U(\mathcal S) +\frac{1}{2}\sum_{l=1}^k l(3-l)\sum_{T\in S_l}p'_T\\
	       & = U(\mathcal S) -\sum_{l=1}^k \sum_{T\in S_l}p'_T +\frac{1}{2}\sum_{l=1}^k l(3-l)\sum_{T\in S_l}p'_T\\
				& = U(\mathcal S) +\frac{1}{2}\sum_{l=1}^k (-l^2+3l-2)\sum_{T\in S_l}p'_T\\
				& = U(\mathcal S) - \frac{1}{2}\sum_{l=3}^k (l-1)(l-2)\sum_{T\in S_l}p'_T.
\end{align*}
Above we have used the fact that $U(\mathcal S) = \sum_{l=1}^k \sum_{T\in S_l}p'_T$.  We also adjusted the limits of the final sum because the terms $(l-1)(l-2)$ are zero for $l=1,2$.  {Because $(l-1)(l-2)\geq 0$ for all $l \geq3$, we obtain the lower bound $U(\mathcal S)-U_2(\mathcal S)\geq 0$.}

We next upper bound  $U(\mathcal S)-U_2(\mathcal S)$.  Recall that we assumed for any $1\leq l \leq k$ and any $T\in \mathcal S_l$ that \hunter{$p'_T\leq cp^l$}.  {In addition note that $|\mathcal S_l| = {k\choose l}$ for $l\in\curly{0,1,...,k}$ since this is the number of sets of $l$ elements chosen from a set of carnality $k$.  With this assumption we can write the difference as}
\begin{align*}
	U(\mathcal S)-U_2(\mathcal S) & = \frac{1}{2}\sum_{l=3}^k (l-1)(l-2)\sum_{T\in \mathcal S_l}p_T'\\
	                        & \leq \frac{c}{2}\sum_{l=3}^k (l-1)^2 {k\choose l}p^l  \\
	                        &\leq \frac{c}{2}\sum_{l=3}^k (l-1)^2 (kp)^l\\
	                        &\leq \frac{c}{2}\paranth{\sum_{l=2}^k (l-1)^2 (kp)^l  -(kp)^2}\\
	                        &\leq \frac{ckp}{2}\paranth{\sum_{l=2}^k (l-1)^2 (kp)^{l-1}  -kp}\\
                           &\leq \frac{ckp}{2}\paranth{\sum_{l=1}^\infty l^2 (kp)^l-kp}\\
                            &\leq \frac{ckp}{2}\paranth{\frac{kp+(kp)^2}{(1-kp)^3}-kp}\\
                            &\leq \frac{ckp}{2}\paranth{\frac{kp+(kp)^2-kp+3(kp)^2-3(kp)^3+(kp)^4}{(1-kp)^3}}\\
                            &\leq \frac{ckp}{2}\paranth{\frac{4(kp)^2+(kp)^4}{(1-kp)^3}}\\
\end{align*}
{We now use the fact that $kp<1/2$ to lower bound $(1-kp)^3$ by $1/8$ and upper bound $(kp)^4$ by $(kp)^2$.  Using these bounds we obtain the upper bound $U(\mathcal S)-U_2(\mathcal S) \leq 20c(kp)^3.$}
\subsection{Proof of Lemma \ref{lem:U2_pt}}
We assume without loss of generality that \hunter{$\mathcal S = \curly{E_i}_{i=1}^k$}.
Then we can write $U_2(\mathcal S)$ as
\begin{align*}
	U_2(\mathcal S)& = \sum_{i=1}^k\sum_{l=1}^k\sum_{\substack{T\in\mathcal S_l\\ E_i\in T}}p'_T-\frac{1}{2}\sum_{\substack{i,j=1\\j\neq i}}^k\sum_{l=2}^k\sum_{\substack{T\in\mathcal S_l\\E_i,E_j\in T}}p'_T\\
	& = \sum_{l=1}^k {l\choose 1}  \sum_{T\in\mathcal S_l}p'_T-\sum_{l=2}^k{l\choose 2} \sum_{T\in\mathcal S_l}p'_T\\
	& = \frac{1}{2}\sum_{l=1}^k l(3-l)  \sum_{T\in\mathcal S_l}p'_T.
\end{align*}
Above we have used the convention ${l\choose k} = 0$ for $l<k$ and the relation
$
	{l\choose 1}-{l\choose 2}  =l(3-l)/2$.
\subsection{Proof of Theorem \ref{lem:lower_bound}}
We define the marginal mean and variance of $X_i$ as $\mu_i$ and $\sigma^2_i$ and we define the correlation coefficient of $X_i$ and $X_j$ as $\rho_{ij}$.  We define $p_i = \mathbf P(E_i)$ and $p_{ij} = \mathbf P(E_i\bigcap E_j)$.  We will use the following bounds for Gaussian random variables.
\begin{lemma}\citep{gordon1941values}\label{lem:pi}
Let $X$ be a Gaussian random variable with mean $\mu$ and positive standard deviation $\sigma$.  For any value $t>\mu$ let $z=(t-\mu)/\sigma$.  Then we have
\begin{align}
	\frac{\exp{\paranth{-z^2/2}}}{\sqrt{2\pi}\sigma(z+1/z)}   \leq
	\mathbf P(X\geq t) \leq
	\frac{\exp{\paranth{-z^2/2}}}{\sqrt{2\pi}\sigma z}.
\end{align}
\end{lemma}
We will also use the following lemma to obtain an upper bound on the joint probability $p_{ij}$.
\begin{lemma}\label{lem:pij}
Let $(X_1,X_2)$ be a pair of jointly Gaussian random variables with means $\mu_1,\mu_2$, positive standard deviations $\sigma_1, \sigma_2$, and correlation coefficient $\rho_{12}$.  For any value $t>\max(\mu_1,\mu_2)$, we have
\begin{align}
	\mathbf P(\min(X_1,X_2)>t)& \leq \frac{1}{\sqrt{2\pi}(2t-\mu_1-\mu_2)} \exp\paranth{-\frac{(2t-\mu_1-\mu_2)^2}{2\paranth{\sigma_1^2+\sigma_2^2+2\rho_{12}\sigma_1\sigma_2}}}.
\end{align}
\end{lemma}
With these bounds for the relevant probabilities, we can obtain a lower bound on the objective function.  Let $z_i = (t-\mu_i)/\sigma_i$.  Then using Lemmas \ref{lem:pi} and \ref{lem:pij} we have
\begin{align}
	U_2(S) =&\sum_{i:E_i\in \mathcal S}p_i-\frac{1}{2}\sum_{\substack{i,j:\\E_i,E_j\in \mathcal S\\i\neq j}}p_{ij}\nonumber\\
	\geq& \sum_{i:E_i\in \mathcal S} \frac{1}{\sqrt{2\pi}\sigma_i(z_i+1/z_i)}\exp{\paranth{-\frac{z_i^2}{2}}}\nonumber\\
	&-\frac{1}{2}\sum_{\substack{i,j:\\E_i,E_j\in \mathcal S\\i\neq j}} \frac{1}{\sqrt{2\pi}(2t-\mu_1-\mu_2)} \exp\paranth{-\frac{(2t-\mu_i-\mu_j)^2}{2\paranth{\sigma_i^2+\sigma_j^2+2\rho_{ij}\sigma_i\sigma_j}}} \nonumber\\
		& = U^l_2(\mathcal S)\nonumber.
\end{align}
\subsection{Proof of Lemma \ref{lem:pij}}
{
We can upper bound the probability of interest by
\begin{align*}
    	\mathbf P(\min(X_1,X_2)>t)& \leq 	\mathbf P(X_1+X_2 > 2t)
\end{align*}
The sum $X_1+X_2$ is a Gaussian random variable with mean $\mu_1+\mu_2$ and variance $\sigma_1^2+\sigma_2^2+2\rho_{12}\sigma_1\sigma_2$.  By applying Lemma \ref{lem:pi} we obtain
the desired result.}

\section{Additional Formulations}\label{formulation_appendix}

\subsection{Linearizing Binary Quadratic Constraint}

{ To linearize quadratic constraint \eqref{eq:overlaptwo} we just need to introduce binary variables $z_j\in \{0,1\}$ for $j=1,\ldots,p$ that we relate to $x_{(i+1)j}$ and $x_{ij}$ through
\hunter{
\begin{equation}
\begin{aligned}
& & &   z_j\leq x_{ij}, & j=1,\ldots,p,\\
& & &   z_j\leq x_{(i+1)j}, & j=1,\ldots,p,\\
& & &   z_j\geq x_{ij}+ x_{(i+1)j}-1, & j=1,\ldots,p.
\end{aligned}
\end{equation}
}
 Under these constraints, \eqref{eq:overlaptwo} is equivalent to
\begin{equation}
\begin{aligned}
\hunter{$\sum_{j=1}^{p}z_j \leq \gamma$}.\\
\end{aligned}
\end{equation}}
\subsection{Strengthened Goalie and Line Stacking Constraint}
{
A standard way to strengthen Goalie constraint \eqref{eq:no_goalie} is to replace it with the following set of constraints
\hunter{
\begin{equation}\label{eq:no_goalie_st}
\begin{aligned}
& & &  x_{ij} + x_{il} \leq 1, & \forall j \in G, \quad \forall l \in O_j.
\end{aligned}
\end{equation}
}
Similarly, Line Stacking constraints \eqref{eq:complete_line} can be strengthened to
\begin{equation}\label{eq:complete_line_st}
\begin{aligned}
& & \hunter{ $v_{il} $}&\hunter{$\leq  x_{ij},$} &\hunter{$l=1, \ldots, N_L, \; \forall j \in L_l$}&  \\
&&  \hunter{ $\sum\limits_{l =1}^{N_L} v_{il}$} & \hunter{ $\geq 1$} & \\
&&  \hunter{ $v_{il}$} & \hunter{ $\in \left\{0,1 \right\}, $} & \hunter{ $ \hunter{$l=1, \ldots, N_L$}.$}
\end{aligned}
\end{equation}
Unfortunately, this change increases the size of the formulation and modern solvers can often dynamically implement the strengthening as needed \citep[page 69]{conforti2014integer}, which can make it hard to predict the potential improvement of switching \eqref{eq:no_goalie}/\eqref{eq:complete_line} by \eqref{eq:no_goalie_st}/\eqref{eq:complete_line_st}. Figure \ref{fig:runtime_weak_strong} shows that for our instances the switch actually increases solve times.
\begin{figure}\centering
	\includegraphics[scale=.35]{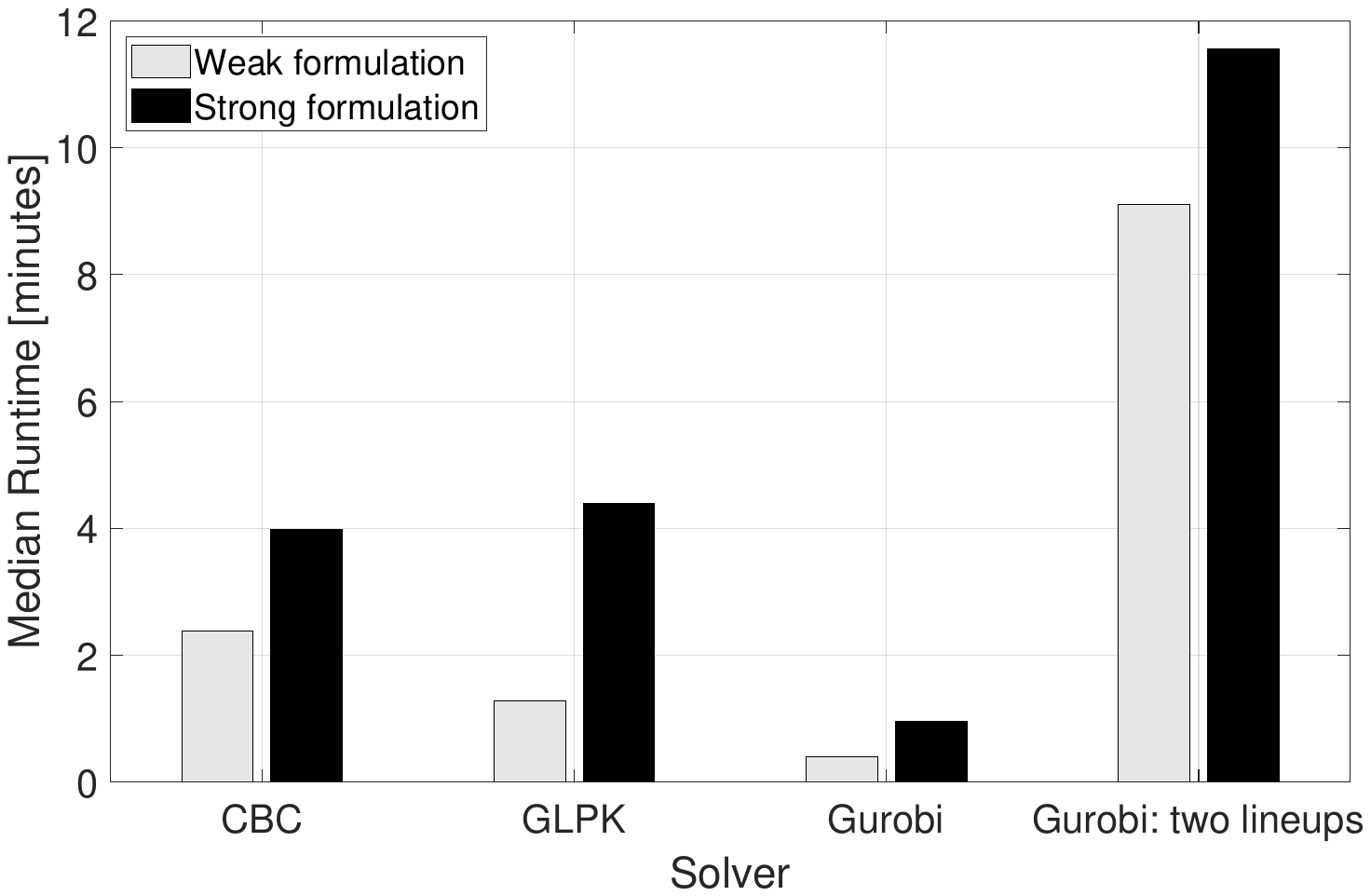}
	\caption{The runtime of our algorithm using weak and strong formulations for generating 100 lineups using Type 4 constraints with an overlap parameter of 6 for 38 different hockey contests with different integer programming software.  The bars labeled ``Gurobi: two lineups'' are the runtimes when solving two lineups at a time using Gurobi.  The other bars are for the sequential approach where we solve for one lineup at a time.  }
	\label{fig:runtime_weak_strong}
\end{figure}
}


\bibliographystyle{plainnat}
\bibliography{DK_bib}
\end{document}